\documentclass{aa}  
\usepackage{graphicx}
\usepackage{amsmath}
\usepackage{natbib}
\usepackage{subfigure}
\usepackage{txfonts}
\usepackage{physics}
\begin{document}
	\nolinenumbers

   \title{The effect of thermal misbalance on magnetohydrodynamic modes in coronal magnetic cylinders}

\author{S.M.Hejazi
	\inst{1,}\inst{2}
	\and
	T. Van Doorsselaere\inst{1}
	\and M. Sadeghi\inst{1}
	\and D. Y. Kolotkov\inst{3,}\inst{4}
	\and J. Hermans\inst{1} 
}

\institute{Centre for mathematical Plasma Astrophysics (CmPA), KU Leuven,
	Celestijnenlaan 200B bus 2400, 3001 Leuven, Belgium\\
	\email{tom.vandoorsselaere@kuleuven.be}
	\and
	Department of Physics, Kharazmi University, 49 Dr. Mofatteh Avenue, Tehran 15719‑14911, Iran\\
	\email{seyedmahmoodhejazi@khu.ac.ir}
	\and
	Centre for Fusion, Space and Astrophysics, Department of Physics, University of Warwick, CV4 7AL Coventry, UK
	\and
	Engineering Research Institute \lq\lq Ventspils International Radio Astronomy Centre (VIRAC)\rq\rq, Ventspils University of Applied Sciences, Ventspils, LV-3601, Latvia\\
}

\date{Received: 15 May 2024 / Accepted: 08 January 2025}

% \abstract{}{}{}{}{}
% 5 {} token are mandatory

\abstract
  
{It is well demonstrated that thermal misbalance, arising from the discrepancy between optically thin radiative energy loss and heating energy gain, disrupts the adiabatic nature of solar corona plasmas, directly affecting the propagation of slow magnetoacoustic waves. However, the extent to which this thermal misbalance, acting as a dispersion factor of an arbitrary intensity, influences the use of slow modes as seismological tools and affects sausage and kink harmonic modes within a magnetic plasma flux tube, remains unresolved.}
{This study investigates the dispersion of magnetohydrodynamic waves influenced by thermal misbalance in a cylindrical configuration with a finite axial magnetic field within solar coronal plasmas. Specifically, it examines how thermal misbalance, characterized by two distinct timescales directly linked to the cooling and heating functions, influences the dispersion relation. This investigation is a key approach for understanding non-adiabatic effects on the behaviour of these waves.}
{The analysis explores the impact of non-adiabatic effects due to classical thermal misbalance, where the heating and cooling timescales vary across a range of values corresponding to each magnetohydrodynamic mode. The dispersion relation for magnetohydrodynamic waves propagating through a magnetic plasma tube, aligned with a finite magnetic field, is calculated under coronal conditions in the linear regime.}
{Our findings reveal that the effect of thermal misbalance on fast sausage and kink modes, consistent with previous studies on slabs, is small but slightly more pronounced than previously thought. The impact is smaller at long-wavelength limits but increases at shorter wavelengths, leading to higher damping rates. This minor effect on fast modes occurs despite the complex interaction of thermal misbalance terms within the dispersion relation, even at low-frequency limits defined by the characteristic timescales. Additionally, a very small amplification is observed, indicating a suppressed damping state for the long-wavelength fundamental fast kink mode. In contrast, slow magnetoacoustic modes are significantly affected by thermal misbalance, with the cusp frequency shifting slightly to lower values, which is significant for smaller longitudinal wavenumbers. This thermal misbalance likely accounts for the substantial attenuation observed in the propagation of slow magnetoacoustic waves within the solar atmosphere. The long-wavelength limit leads to an analytical expression that accurately describes the frequency shifts in slow modes due to misbalance, closely aligning with both numerical and observational results.}


\maketitle

\section{Introduction}

Magnetohydrodynamic (MHD) waves are fundamental to both experimental and astrophysical plasma physics, and many phenomena observed in the solar atmosphere are attributed to their influence. It is well theorized that these waves play a crucial role in the transfer of energy from the lower to the upper layers of the solar atmosphere, offering a key explanation for coronal heating  \citep[see e.g.][]{mcintosh,DeMoortel15,Khomenko,TVD2020}. Moreover, MHD waves serve as natural seismological tools, providing valuable insights into the physical properties of solar plasmas \citep{Ofman14} \citep[for a recent review see e.g.][]{Nakariakov2024}. They are also implicated in the acceleration of solar winds \citep{Cranmer12, Hejazi21}, and their non-linear effects may contribute to the geometry of solar phenomena such as collimation of coronal jets \citep{Hejazi17}.

A key manifestation of MHD waves in the solar atmosphere is through magnetoacoustic (MA) modes, classified by their symmetry properties such as sausage and kink waves, which manifest in both fast and slow modes as fundamental examples of MHD wave phenomena. These modes are characterized by distinct oscillation patterns, frequencies, and phase speeds along magnetic structures, and have been widely observed in various solar structures, including coronal loops, jets, and prominences \citep{Nakariakov99, TVD2008, Dorotovic, DeMoortel2015, Moreels15, Nakariakov2020}. The deviation from thermal equilibrium caused by MHD waves, leading to enhanced plasma heating over cooling in the upper layers of the solar atmosphere through various damping mechanisms, has been extensively studied \citep[for a comprehensive review, see][]{TVD2020}. Conversely, the heating can influence the dynamics of  MHD waves, detectable through a broad range of spectroscopic observations, from microwave to X-ray emission lines \citep{Kolotkov2019}. The observation of MA waves in the solar atmosphere serves as an effective seismological tool, where the wave properties —such as propagation speeds, periods, amplitudes, and damping times— enable the determination of essential plasma parameters, including density, temperature, magnetic field strength, and heating function. Several theoretical and observational studies have highlighted the potential of MHD seismology to reveal physical properties of solar atmospheric plasmas \citep{Andries2009, Verwichte, Kolotkov2020, Nakariakov2021}.

Slow MA waves, whether propagating or standing, are fundamental components of the MHD wave spectrum observed and theorized in the solar atmosphere. These waves are detected via Doppler shifts or intensity oscillations in microwave, extreme-ultraviolet (EUV), and X-ray emission bands \citep{NakaSlow2000, Nakariakov2019, Wang2021}. Due to their ubiquitous presence and relative ease of detection compared to other MHD waves, they have attracted considerable attention as diagnostic tools for probing solar plasma characteristics \citep{Wang2003, Nakariakov2005, DeMoortel2009}. One notable example of slow MA waves is the so-called SUMER oscillations, named after observations by the Solar Ultraviolet Measurements of Emitted Radiation (SUMER) instrument on board the Solar and Heliospheric Observatory (SoHO) \citep{Wang2002, Wang2011}. These oscillations, also often observed with instruments such as the Interface Region Imaging Spectrograph (IRIS) \citep{Li2017}, manifest as rapidly damped, standing, slow-mode oscillations. They are commonly characterized by periodic or quasi-periodic fluctuations in temperature and density within hot coronal loops, providing crucial insights into the physical processes governing solar dynamics.

A particularly interesting aspect of wave behaviour that can significantly influence the propagation characteristics of MA modes, including their damping and amplification within magnetic flux tubes, lies in the role of thermal misbalance, a non-adiabatic process resulting from the disparity between the timescales of heating and radiative cooling perturbations in the plasma. The phenomenon of thermal instability in plasma, first studied by \citet{Field}, revealed that thermal equilibrium can be inherently unstable in a uniform medium, leading to the formation of condensations of a higher density and lower temperature. This study provided foundation under how thermal instability operates under a wide range of conditions, influencing not only solar phenomena but also broader astrophysical structures such as planetary nebulae and interstellar mediums. Earlier studies on thermal misbalance, such as those done by \citet{Nakariakov2000}, \citet{Kumar}, \citet{Kolotkov2019}, and \citet{Zavershinskii}, further developed the understanding of wave-induced thermal misbalance under the approximations of infinite magnetic field and weak non-adiabaticity, highlighting its effects on the evolution of slow MA waves.
	
Recent studies have shown that slow MA waves, particularly in solar coronal loops, are strongly influenced by thermal misbalance. This misbalance causes wave amplification, attenuation, and dispersion, depending on the dominant thermodynamic processes in the plasma. Observations have shown that slow MA waves undergo frequency-dependent damping, which traditional damping theories fail to explain. This opens the possibility for using thermal misbalance to diagnose unknown coronal heating mechanisms through detailed seismological analysis of wave properties \citep{Nakariakov2017, Kolotkov2019, 2022MNRAS.514L..51K}. In both propagating and standing MA waves, heating-cooling misbalance can result in different damping regimes, oscillatory patterns, and even wave amplification, depending on the plasma conditions \citep{Zavershinskii,2020SoPh..295..160B, Kolotkov2020, Agapova}. This has led to ongoing efforts to better understand how wave properties such as phase shifts, group speeds, and damping lengths are affected by both the plasma's non-adiabatic processes and its magnetic structuring \citep{Dmitrii2021, 2022SoPh..297....5P, 2022SoPh..297..144I}.
	
However, the solar corona is a complex environment where the infinite magnetic field approximation is not universally applicable. Variations in magnetic field strength, as reported in the works by \citet{Afanasyev} and \citet{Nakariakov2017}, suggest that the assumption of an overwhelmingly dominant magnetic pressure may not always hold. Thus, it becomes crucial to explore scenarios where the magnetic field strength is finite, corresponding to non-zero plasma-$\beta$ conditions, and to assess their impact on wave dynamics. This consideration is particularly relevant since coronal heating, believed to be primarily driven by magnetic fields, can have its efficiency modulated by the field strength itself, consequently affecting the thermal misbalance and the associated behaviour of MHD waves \citep[e.g.][]{Duckenfield}. Building on this, \citet{Kolotkov2019, Kolotkov2020} presented compelling evidence of rapidly decaying slow MA waves due to thermal misbalance observed in hot coronal loops, demonstrating its potential for constraining the coronal heating function through observed damping rates of slow modes. These studies further conclude that coronal plasmas, in all scenarios, should always be regarded as an active medium, powered by thermal misbalance and engaged in a continuous exchange of energy with MHD waves \citep{Dmitrii2021}.
	
In addition to affecting slow MA waves, thermal misbalance also influences other MHD wave modes, including Alfv\'en waves. These waves are subject to non-linear effects due to the interaction of torsional and shear Alfv\'en waves with the surrounding plasma. Studies have shown that the induced longitudinal velocity perturbations in such waves are modulated by thermal misbalance, with the effect being more pronounced in shear waves than torsional ones \citep{10.1093/mnras/stac2066}. Furthermore, the back-reaction caused by the perturbed thermal equilibrium in the solar corona results in thermal instabilities, which play a key role in the formation of fine thermal structures in coronal loops \citep{Dmitrii2023}. These instabilities are stabilized to some extent by thermal conduction, especially in hot active regions, but the sensitivity of slow wave stability to coronal heating functions presents an opportunity for further seismological diagnostics \citep{2023FrASS..1067781Z, 2024RSPTA.38230222B}. Overall, this demonstrates the effect of thermal misbalance on MHD wave dynamics and highlight its importance in constraining coronal heating mechanisms and explaining solar atmospheric behaviour \citep{2022MNRAS.514L..51K,2023MNRAS.522..572R}.

The implications of these findings are significant, potentially resolving long-standing discrepancies in the observed frequency-dependent damping of slow MA waves in various coronal structures, as reported by \citet{Mariska}, \citet{Gupta}, \citet{Krishna2014}, \citet{Krishna2018}, and \citet{2022MNRAS.514L..51K}. Numerical studies have demonstrated that damping lengths vary with wave periods \citep{Gupta, Mandal}. Furthermore, \citet{Arregui2023} highlighted the importance of considering both thermal conduction and thermal misbalance in explaining the damping of slow MA waves in hot coronal loops. The consideration of thermal misbalance as a damping mechanism may reconcile observed phase shifts between density and temperature, as well as the growth of the poly-tropic index, with the theoretical predictions done by \citet{Owen}, \citet{VanDoorsselaere}, and \citet{Krishna2018}. Additionally, \citet{Dmitrii2022} explored the use of slow wave seismology to compare the effects of weak and strong thermal conduction, aiming to derive a more accurate value for the effective adiabatic index in coronal plasma.

In this study, we develop an analytical model to investigate the impact of thermal misbalance in none zero plasma-$\beta$ on a spectrum of MHD modes in coronal magnetic structures, using coronal seismology as a tool to refine the understanding of the heating function. The MHD modes analysed in this research cover a range of phase speeds, from fast to slow MA waves, propagating within solar magnetic cylinders. By expanding the non-adiabatic linear dispersion relation to the long wavelength limit, we aim to provide deeper insight into the damping times of slow MA waves in the solar corona and the dependence of wave damping on the heating and cooling functions. The context of the present study and its associated model are inspired by a series of works by \citet{EandR}, \citet{VDL91}, \cite{Ireland92}, and \citet{Kolotkov2019}, to advance the field by incorporating the geometrical constraints of a cylindrical configuration with consideration of finite magnetic field, resulting transversal structuring, alongside the influence of thermal misbalance of arbitrary intensity, which are intrinsic to the solar coronal environment.

%__________________________________________________________________

\section{Governing equations and dispersion relation}

We consider a magnetic cylinder in a uniform medium characterized by equilibrium finite magnetic field $\vec{B}_{_0}=(0,0,B_{_0})$, pressure $p_{_0}$, and density $\rho_{_0}$. Our analysis employs the ideal MHD equations while also considering the influence of thermal misbalance between heating and optically thin cooling radiation:

\begin{align}
	&\rho [\frac{\partial {\bf v}}{\partial t} +({\bf v}. \nabla){\bf v}] = -\nabla p + \frac{1}{4\pi} (\nabla \times \vec{B}) \times \vec{B},  \label{eq:1}\\
	&\frac{\partial \rho}{\partial t} + \nabla .(\rho {\bf v}) = 0,   \label{eq:2}\\
	&\frac{\partial \vec{B}}{\partial t} = \nabla \times ({\bf v} \times \vec{B}),   \label{eq:3}\\
	&p = \frac{k_{_{B}} \rho T}{m},   \label{eq:4}\\
	&\frac{\rho^\gamma}{\gamma -1} \frac{d}{dt} (\frac{p}{\rho^\gamma}) = - \rho \emph{Q}(\rho,T),   \label{eq:5}\\
	&\nabla\cdot\vec{B} = 0,   \label{eq:06}
\end{align}
where, $\emph{Q}(\rho,T)$ represents the discrepancy between the heating and the radiative cooling processes, defined as $L(\rho,T) - H(\rho,T)$ \citep{Parker}. The cooling function, $L(\rho,T)$, and the heating function, $H(\rho,T)$, are both expressed as general power-law functions of density and temperature as below, with unspecified exponents, each of which must be independently determined for different regions within the solar atmosphere,

\begin{equation}
	L(\rho,T)= \chi \rho T^{\beta}, \hspace*{+1cm} H(\rho,T)= \emph{h} \rho^{a} T^{b}.   \label{eq:6}
\end{equation}

The optically thin radiative cooling function, $L(\rho,T)$, is readily determinable, thanks to the CHIANTI atomic database \citep[since][]{Dere97}. In this study, we employed the latest version, 10.1 \citep{Dere2023}, to calculate the temperature exponent, $\beta$, and the coefficient $\chi$ in the cooling function. Regarding the unspecified heating function, it is necessary to consider the constraints governing the values of the power-law exponents $a$ and $b$ \citep{Kolotkov2020}. Assuming an isothermal initial equilibrium, we set $\emph{Q}(\rho_{_0}, T_{_0}) = 0$, which is essential for determining the coefficient of $\emph{h}$ in the heating function. It is noteworthy that some heating mechanisms may exhibit time-varying components on very short timescales \citep{Reale} or dependence on the magnetic field \citep{Duckenfield}. In this study, we adopt a non-specified model for both heating and cooling functions. Our primary objective is to develop a descriptive model to understand how arbitrary forms of thermal misbalance functions, each characterized by its own timescales, influence MHD modes within a magnetic cylindrical waveguide. While this approach may lead to precise forms of the heating function, such details are beyond the scope of this article.

At this point, it is important to note that both the general heating and cooling functions depend on density, and the heating coefficient, $\emph{h}$, is directly linked to the background density profile, $\rho_{0}$, where under realistic solar atmospheric conditions, should be considered stratified. Consequently, assuming a uniform background plasma may be inaccurate, and the effects of stratification should be incorporated into our mathematical models. However, it is essential to recognize that the MA waves, particularly slow MA waves considered in this study, are characterized by compressive oscillations and represent the MHD modes significantly influenced by thermal non-adiabaticity in the solar corona \citep{Zavershinskii, Kolotkov2019, Kolotkov2020, Agapova}. These waves exhibit wavelengths ranging from a few thousand to several tens of thousands of kilometres. Notably, this range is much smaller than the length scale of stratification, which reaches a minimum value of approximately 50 Mega-metres \citep{NakaOfman2000, Andries2005}. Therefore, the assumption of a uniform medium remains justified.

\begin{figure}
	\centering\includegraphics[width=0.8\linewidth]{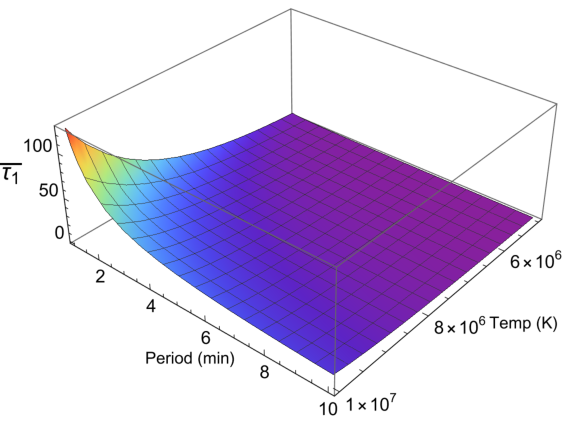}
	\centering\small (a)
	\hfil
	\centering\includegraphics[width=0.8\linewidth]{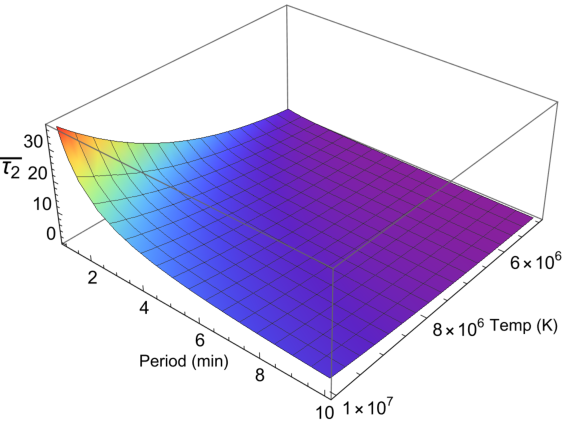}
	\centering\small (b)
	
	\caption{Dependence of dimensionless characteristic timescales, $\bar{\tau}_1$ (a) and $\bar{\tau}_2$ (b), on temperature and wave period calculated using Eq.~(\ref{eq:10}), normalized by the wave period. This analysis considers a heating model characterized by power indices $a=-0.5$ and $b=-3$ for density and temperature, respectively.}
	\label{figure2}
\end{figure}

Upon linearizing the governing equations by introducing small amplitude perturbations to each physical parameter around the initial equilibrium, we focus on the variable that represents compressibility in the system ($\nabla \cdot {\bf v}$) \citep{EandR}. This procedure results in a linear set of operators that govern the physical characteristics of the system, as outlined below

\begin{equation}
	(\widehat{G}+\widehat{\emph{Q}})\Delta = 0.   \label{eq:7}
\end{equation}
Let $\Delta = \nabla\cdot{\bf v}$, and the subsequent expressions represent each operator:

\begin{equation}
	\begin{array}{l}
		\widehat{G} \equiv \frac{\partial^2}{\partial t^2} \left[\frac{\partial^2}{\partial t^2}-(V_{A}^2+C_{s}^2) \nabla^2 \right]+ V_{A}^2 C_{s}^2 \frac{\partial^2}{\partial z^2} \nabla^2,\\
		\widehat{\emph{Q}} \equiv \emph{Q}_{\rho} \left[\rho_{_0} (\gamma-1) \frac{\partial}{\partial t} \nabla^2\right]\\ \qquad+ \emph{Q}_{_{T}} \left[\frac{1}{C_{_V}}\frac{\partial}{\partial t}\left(\frac{\partial^2}{\partial t^2}-V_{A}^2 \nabla^2+ V_{A}^2 \frac{\partial^2}{\partial z^2}\right)-(\gamma-1) T_{_0} \frac{\partial}{\partial t} \nabla^2\right]. \label{eq:8}
	\end{array}
\end{equation}

Each differential operator in Eq.~(\ref{eq:7}) corresponds to a specific physical effect within the system. The operator $\widehat{G}$ represents the geometrical constraints accounted for in the dispersion relation, which can incorporate the cylindrical magnetic configuration \citep[see also Eq.~(3a) in][]{EandR}. The operator $\widehat{\emph{Q}}$ describes the effect of thermal misbalance between heating and radiative cooling processes as well. For further details on the derivation of the mathematical expressions of the operators in Eq.~(\ref{eq:8}), refer to Appendix \ref{appendix:A}.

In Eq.~(\ref{eq:8}), $V_{A}$ and $C_s$ represent the Alfv\'en and acoustic speeds in the system, respectively. They are defined by the relations $V_{A} = B_{_0} (4\pi \rho_{0})^{-1/2}$ and $C_s = \sqrt{\gamma k_{_{B}} T_{_0} m^{-1}}$, where $B_{_0}$ is the equilibrium magnetic field strength, $\rho_{_{0}}$ is the background density, $\gamma$ is the adiabatic index, $k_{_{B}}$ is the Boltzmann constant, $T_{_0}$ is the equilibrium temperature, and $m$ is the mean particle mass. Additionally, the specific heat capacity at constant volume, denoted as $C_{_V} = (\gamma - 1)^{-1} k_{_{B}}/m$, is determined using the standard adiabatic index ($\gamma = 5/3$). It is important to note that the polytropic index of the system deviates from the standard adiabatic index when non-adiabatic terms are included in the energy equation (Eq.~(\ref{eq:5})). Adjusting this index becomes necessary to account for the effects on the index in subsequent calculations. However, the focus of this analytical study is on the complex influence of thermal misbalance on MHD waves, and therefore, this topic is reserved for future investigation. For more detailed discussions on the effective adiabatic index in coronal plasma, readers are encouraged to refer to \citet{VanDoorsselaere}, \citet{Krishna2018}, and \citet{Dmitrii2022}.

Moreover, in  Eq.~(\ref{eq:8}), We express the rate of change of the heating-cooling function with respect to density at constant temperature as $\emph{Q}_{\rho}$, and with respect to temperature at constant density as $\emph{Q}_{T}$, as follows:

\begin{equation}
	\emph{Q}_{\rho}\equiv \frac{\partial \emph{Q}}{\partial \rho}|_{_{T}} \hspace*{+.1cm},\hspace*{+.7cm} \emph{Q}_{_{T}}\equiv \frac{\partial \emph{Q}}{\partial T}|_{\rho}\hspace*{+.2cm}.   \label{eq:9}
\end{equation}

In the thin flux tube approximation, neglecting parallel thermal conduction reduces Eq.~(10) in \citet{Dmitrii2021} to Eq.~(\ref{eq:7}). It is also deducible that neglecting the effect of thermal misbalance in Eq.~(\ref{eq:7}), results in the presence of only the geometrical dispersion operator, reducing it to Eq.~(3a) in \citet{EandR}. At this point, it is noteworthy that if we consider the governing equations in the absence of magnetic effects (i.e. in the hydrodynamic limit) while accounting for thermal conduction, the linearized perturbed equations (Eqs.~(\ref{eq:32}, \ref{eq:33}, \ref{eq:35}, \ref{eq:36}) in Appendix \ref{appendix:A}) regenerate the forms of Eqs.~(11-14) in \citet{Field}. This ends up resulting in the dispersion relation given by Eq.~(15) in \citet{Field}, for an exponentially propagating wave solution in an infinite, uniform medium. 

By defining the characteristic timescales $\tau_1$ and $\tau_2$ as follows, it becomes possible to quantify the effects of cooling and heating processes on MA waves, for instance, regarding attenuation, amplification, and phase speed changes \citep{Zavershinskii},

\begin{equation}
	\tau_1 = \gamma C_{_V}/[\emph{Q}_{T}-\frac{\rho_{_0}}{T_{_0}}\emph{Q}_{\rho}]\hspace*{+.1cm},\hspace*{+.7cm}\tau_2=C_{_V}/\emph{Q}_T.   \label{eq:10}
\end{equation}

Additionally, in the infinite magnetic field approximation—where the background magnetic field is considered infinitely strong, allowing the perturbation of the magnetic fields in the system to be assumed negligible —the geometrical dispersion effect on slow modes can be disregarded. In this scenario, the governing equations are reduced to one-dimensional forms along the $\hat{z}$ axis. Therefore, incorporating the characteristic timescale for thermal conduction, $\tau_{\text{cond}}$, which is proportional to the square of wavelength \citep[see][]{Zavershinskii}, allows for the reduction of Eq.~(\ref{eq:7}) to Eq.~(5) in \citet{Kolotkov2019}.

In this study, we focus on a segment of the magnetic flux tube located within the coronal plasma, where the $\tau_{\text{cond}}$ is significantly longer than the characteristic timescales associated with thermal misbalance \citep{Kolotkov2019}. Moreover, due to the location in the coronal region of the solar atmosphere, we assume a uniform plasma temperature aligned with the background magnetic field, with a negligible background temperature gradient. This assumption, widely adopted in theoretical coronal plasma studies \citep[e.g.][]{Owen,Mandal,Wang18,Kolotkov2020}, implies that, compared to thermal misbalance, the effects of thermal conduction can be neglected. This allows us to focus exclusively on the influence of thermal misbalance on wave damping \citep[pure thermal misbalance case in][]{Kolotkov2019}. 

In this context, it is where to introduce two dimensionless characteristic timescales, obtained by normalizing $\tau_1$ and $\tau_2$ by the wave period ($P = 2\pi/\omega_{r},\hspace*{+.1cm}\omega_{r}=\mathfrak{Re}(\omega)$), to account for the thermal misbalance. These are defined as $\bar{\tau_1} = \frac{\tau_1}{P}$ and $\bar{\tau_2} = \frac{\tau_2}{P}$. The dependence of these characteristic timescales on the wave period is significant, especially in scenarios where thermal misbalance perturbation exhibits a wave-induced nature \citep{Zavershinskii, Kolotkov2019, Belov}. Through introducing these two dimensionless period-dependent timescales, the impact of thermal misbalance on a wave, characterized by physical parameters such as frequency or period, can be directly correlated with the wave’s susceptibility to thermal misbalance. 

Figure~\ref{figure2} illustrates the dependence of $\bar{\tau_1}$ and $\bar{\tau_2}$ on temperature and wave period. The cooling function is derived from the CHIANTI database v. 10.1 for an equilibrium density ($\rho{_{0}}$) of $2.3\times10^{-15} g/cm^3$, while the heating function is parametrized with power-law exponents $a = -0.5$ and $b = -3$. It is evident that these characteristic timescales satisfy the conditions $\bar{\tau_2} < \bar{\tau_1}$ and $\bar{\tau_1} - \bar{\tau_2} > 0$, ensuring that thermal misbalance acts as a damping factor on MHD waves \citep{Kolotkov2020}. Furthermore, the difference between the equilibrium densities of the external and internal media may cause variations in the characteristic timescales of thermal misbalance. However, this study assumes that such discrepancies in equilibrium densities do not significantly affect the characteristic timescales of thermal misbalance within and outside the tube. Moreover, for waves with longer periods and environmental temperatures ranging from 1 to 5 MK, as shown in Fig.~\ref{figure2}, the values of $\bar{\tau_1}$ and $\bar{\tau_2}$ are minimized, and vice versa.

Next, expressing the Laplacian operator in cylindrical coordinates $(r,\theta,z)$ within the set of differential operators in Eq.~(\ref{eq:8}) as 

\begin{equation}
	\nabla^2 \equiv \frac{\partial^2}{\partial r^2} + \frac{1}{r} \frac{\partial}{\partial r} + \frac{1}{r^2} \frac{\partial^2}{\partial \theta^2} + \frac{\partial^2}{\partial z^2}.   \label{eq:11}
\end{equation}
We adopt the Fourier exponential form to describe the solution’s dependence on the coordinates and time as:

\begin{equation}
	R(r) \exp(i k z + i m \theta - i \omega t),   \label{eq:11.5}
\end{equation}
where $k$, $m$, and $\omega$ represent the longitudinal wavenumber, azimuthal wavenumber, and wave frequency, respectively. Substituting this form into Eq.~(\ref{eq:7}) leads to a Bessel equation for $R(r)$ as

\begin{equation}
	\dv[2]{R}{r} + \frac{1}{r} \dv{R}{r} - (\kappaup^2 + \frac{m^2}{r^2})R = 0\hspace*{+.1cm}.   \label{eq:12}
\end{equation}
Here, $\kappaup$ represents the interior radial wavenumber and is a complex function of wave frequency ($\omega$) derived from the following expression

\begin{equation}
	\kappaup^2 = \frac{(k^2 C_{s}^2 - \omega^2)(k^2 V_{A}^2 - \omega^2) - \frac{i\omega^2}{2 \pi \bar{\tau_2}}(\frac{\bar{\tau_2}}{\bar{\tau_1}} k^2 C_{s}^2 - \omega^2)}{(k^2 C_{T}^2 - \omega^2)(C_{s}^2 + V_{A}^2) - \frac{i\omega^2}{2 \pi \gamma \bar{\tau_2}}(C_{s}^2 + \gamma V_{A}^2)}\hspace*{+.1cm}.  \label{eq:13}
\end{equation}

The tube speed, denoted as $C_{T}$ and expressed as $\sqrt{\frac{C_s^2 V_{A}^2}{C_s^2 + V_{A}^2}}$, is commonly referred to as the cusp speed ($V_{_{C}}$) \citep{Roberts}. However, this study demonstrates that the cusp speed deviates from the tube speed due to the influence of non-adiabatic effects within the system, corroborating earlier findings illustrated in Figs.~(1) and (3) of \citet{Belov}, where it is also referred to as $C_{TQ}$.

Assuming no energy exchange between the interior and exterior of the magnetic flux tube implies that the oscillations within the tube at the boundary are effectively isolated from external energy sources or sinks. This isolation allows the oscillations inside to be treated as independent of external drivers or damping effects. Consequently, we also arrive at a Bessel equation with distinct Alfv\'en, sonic, and tube speeds for the exterior region. The solutions to these Bessel equations, while maintaining continuity at $r = R$ (the radius of the tube) and ensuring finite solutions at both $r = 0$ and $r \rightarrow \infty$, are then obtained by

\begin{equation}\label{eq:14}
	\begin{split}
		R(r) =
		\begin{cases}
			A_{in} \hspace*{+.1cm} J_{m}(\kappaup_{_{in}} r) \hspace*{+1cm} r<R\hspace*{+.1cm},\\
			\\
			A_{out} \hspace*{+.1cm} K_{m}(\kappaup_{_{out}} r) \hspace*{+.8cm} r>R\hspace*{+.1cm},
		\end{cases}
	\end{split}
\end{equation}
where $A_{in}$ and $A_{out}$ are complex constants, and under the condition that $\kappaup_{in}^2=-\kappaup^2>0$ \citep{EandR}, $J_{m}$ and $K_{m}$ are the Bessel function of the first kind and the modified Bessel function of the second kind, respectively, and $\kappaup_{_{out}}$ is the external radial wavenumber, given by

\begin{equation}
	\kappaup_{out}^2 = \frac{(k^2 C_{se}^2 - \omega^2)(k^2 V_{Ae}^2 - \omega^2) - \frac{i\omega^2}{2 \pi \bar{\tau_2}}(\frac{\bar{\tau_2}}{\bar{\tau_1}} k^2 C_{se}^2 - \omega^2)}{(k^2 C_{Te}^2 - \omega^2)(C_{se}^2 + V_{Ae}^2) - \frac{i\omega^2}{2 \pi \gamma \bar{\tau_2}}(C_{se}^2 + \gamma V_{Ae}^2)}\hspace*{+.1cm},   \label{eq:15}
\end{equation}
where $C_{Te} = \sqrt {\frac{C_{se}^2 V_{Ae}^2}{C_{se}^2 + V_{Ae}^2}}$ represents the exterior tube speed. It is worth noting that the radial wavenumbers in Eqs.~(\ref{eq:13}, \ref{eq:15}), as the arguments of the Bessel functions, are complex expressions. therefore, the real part of both Bessel functions $Y_{m}$ and $K_{m}$ meet the condition for finite exterior solutions as the distance from the flux tube boundary increases. However, we take the Bessel function $K_{m}$ as the solution for the external region to remain consistent with the adiabatic limit. This limit corresponds to the characteristic timescales $\tau_1$ and $\tau_2$ approaching infinity, thus recovering the exterior solution in Eq.~(7) of \citet{EandR}. The dispersion relation is then obtained by examining the continuity of the radial velocity component and the total pressure at the tube boundary ($r = R$), expressed as the following function:

\begin{equation}
	D_{m}(\omega) = (k^2 V_{A}^2 - \omega^2) -\frac{\rho_{e}}{\rho_{_0} } \frac{\kappaup_{in}}{\kappaup_{out}} (k^2 V_{Ae}^2 - \omega^2) \frac{J_{m}^{'}(\kappaup_{in} R) K_{m}(\kappaup_{out} R)}{J_{m}(\kappaup_{in} R) K_{m}^{'}(\kappaup_{out} R)} = 0\hspace*{+.1cm}.  \label{eq:16}
\end{equation}

This equation recovers Eq.~(8b) of \citet{EandR}, in the adiabatic limit. As evident in Eq.~(\ref{eq:16}), the distinction between surface and body modes within the tube is no longer clear, as the radial wavenumbers $\kappaup_{_{in}}$ and $\kappaup_{_{out}}$ are complex numbers. This is leading to an intertwined state of surface and body waves, where the solution to the Bessel equation inside the tube ($J_{m}(\kappaup_{in} r)$) becomes a combination of both modes.

Building on the findings of \citet{Agapova} concerning the dynamics of fast and slow MA waves in plasma slabs affected by thermal misbalance, and also supported by the results presented in this study, it becomes obvious that the responses to the dispersion relation are primarily dominated by body modes rather than surface modes.

\section{Long wavelength limit}

For the case $\frac{r}{\lambda} \ll 1$ or, in other words, for small values of the longitudinal wavenumber  $k$, we can implement approximations for Bessel functions with the small arguments as discussed in Section 2. This allows us to derive an analytical expression that describes the effect of thermal misbalance on the MA modes within the flux tube. In this section, mathematical expressions are calculated for the axisymmetric (sausage) and the first azimuthal harmonic (kink) modes.

It is important to note that all the numerical and analytical solutions presented in this section are carried out under typical coronal conditions, where the interior Alfv\'en speed ($V_{Ai}$) is assumed to be 1 Mm/s, the exterior Alfv\'en speed ($V_{Ae}$) is 1.5 Mm/s, the interior sound speed ($C_{si}$) is 0.15 Mm/s, and the exterior sound speed ($C_{se}$) is 0.13 Mm/s. These parameter values refer to the coronal conditions used in all figures throughout this work.

\subsection{Fast sausage modes}
In the case of zero azimuthal wavenumber ($m=0$), the dispersion relation (Eq.~(\ref{eq:16})) takes the following form:

\begin{equation}
	D_{0}(\omega) = (k^2 V_{A}^2 - \omega^2) -\frac{\rho_{e}}{\rho_{_0} } \frac{\kappaup_{in}}{\kappaup_{out}} (k^2 V_{Ae}^2 - \omega^2) \frac{J_{1}(\kappaup_{in} R) K_{0}(\kappaup_{out} R)}{J_{0}(\kappaup_{in} R) K_{1}(\kappaup_{out} R)} = 0\hspace*{+.05cm}.   \label{eq:17}
\end{equation}

\begin{figure}
	\centering\includegraphics[width=0.7\linewidth]{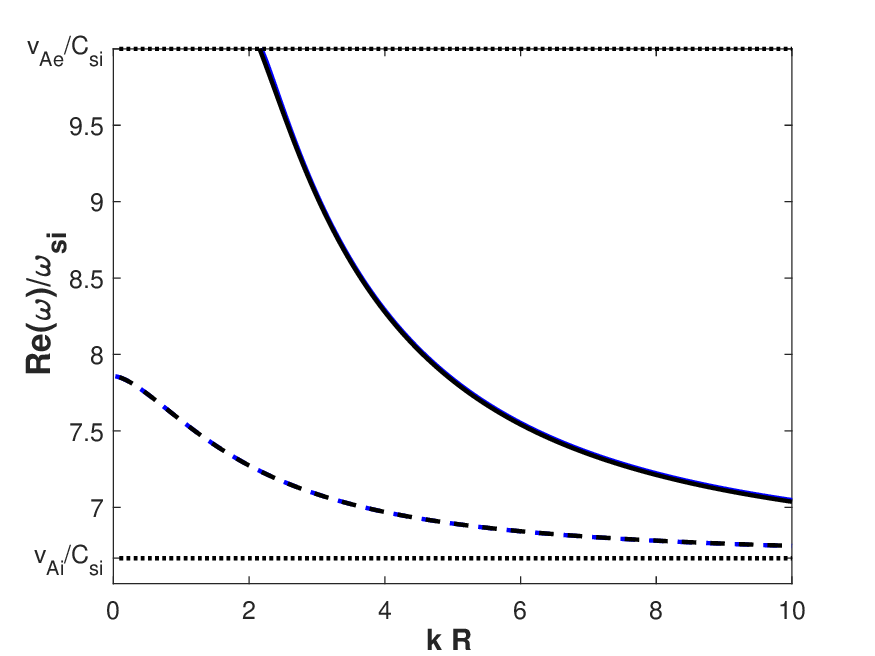}
	\centering\small (a)
	\hfil
	\centering\includegraphics[width=0.7\linewidth]{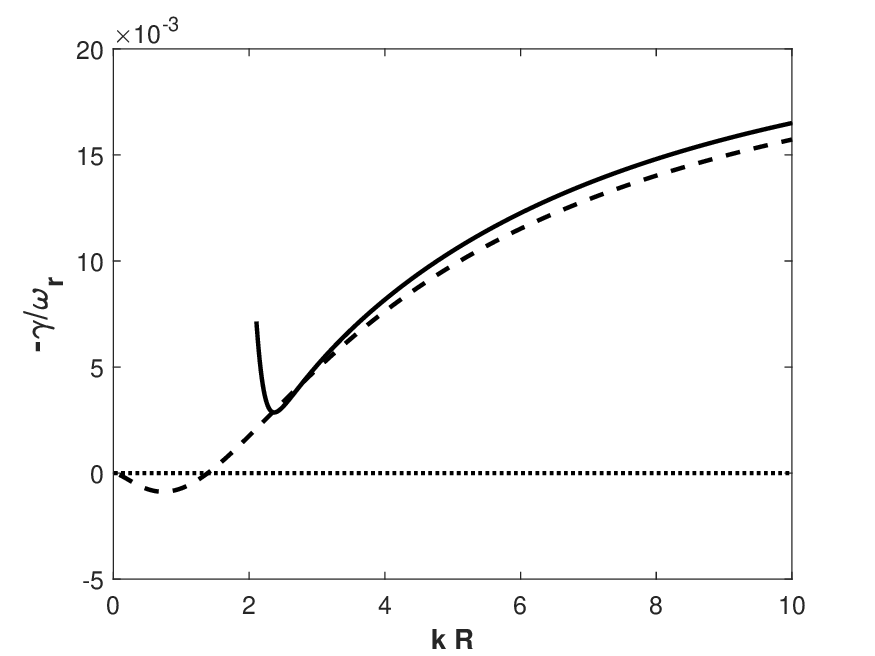}
	\centering\small (b)
	
	\caption{In the coronal condition ($V_{Ae}=1.5, V_{Ai}=1, C_{si}=0.15, C_{se}=0.13$ Mm/s) and with $\bar{\tau_1}=6.8$, $\bar{\tau_2}=2.4$ (corresponding to approximate values of 34 and 12 minutes for $\tau_1$ and $\tau_2$, respectively, at a temperature of about 5 MK), the study considers a wave with a period of 5 min: (a) numerical solution of the dispersion relation from Eq.~(\ref{eq:17}) for frequencies (phase speed) normalized by the internal acoustic frequency (acoustic speed) ($\omega_{r}/\omega_{si} \equiv v_{_{ph}}/C_{si}$) of the fast sausage fundamental mode without (solid blue line) and with (solid black line) consideration of the thermal misbalance effect, as well as the numerical solution of the dispersion relation from Eq.~(\ref{eq:31}) for the fast kink fundamental mode (m=1) without (dashed blue line) and with (dashed black line) consideration of the thermal misbalance effect versus the normalized longitudinal wavenumber ($k R$). (b) Damping rate to frequency ratio ($-\gamma/\omega_{r}$) of fundamental states of the fast sausage (solid line) and kink (dashed line) modes with consideration of the thermal misbalance effect versus the normalized longitudinal wavenumber ($k R$).}
	\label{figure3}
\end{figure}

Figure~\ref{figure3} presents a numerical solution computed using MATLAB to find the roots of the dispersion relation given by Eq.~(\ref{eq:17}). The plot depicts the fundamental MHD modes with phase speeds faster than the internal Alfv\'en speed, which are classified as fast MHD modes. As depicted in Fig.~\ref{figure3}(a) (solid line), the fundamental fast sausage wave exhibits a cut-off in wavenumbers at high frequencies with values exceeding the external Alfv\'en frequency. This cut-off behaviour is an inevitable characteristic of all fast sausage and fast non-fundamental kink modes in the coronal region \citep{EandR}. Consequently, for fast MHD mode responses with frequencies or phase speeds surpassing those of the external Alfv\'en's, the solutions to the Bessel equation in the exterior region manifest in cylindrical Hankel function forms rather than Bessel $K$-function forms. This observation indicates that the waves would exhibit leaky forms of propagation, and the dispersion relation Eq.~(\ref{eq:16}) loses its validity in describing fast MHD modes with phase speeds exceeding the external Alfv\'en speed \citep[for an extensive discussion, refer to][]{Farahani}. Furthermore, in light of a recent study into the cut-off wavenumber observed in the context of sausage oscillations within a slab configuration with varying external density, it is advisable to consult \citet{Wang2023} as a supplementary reference.  

Hence, in this study, it is crucial to confine our analysis to responses where the phase speed does not surpass the external Alfv\'en speed to avoid compromising the validity of the dispersion relation, Eq.~(\ref{eq:16}). Subsequent research on the effect of thermal misbalance on leaky wave solutions could be pursued in future research.

In the regime of long-wavelength sausage modes where the argument of the Bessel functions is approaching to small values, the dispersion relation Eq.~(\ref{eq:17}) takes on an approximate form as

\begin{equation}
	\frac{\rho_{_0} (k^2 V_{A}^2 - \omega^2)}{\rho_{e} (k^2 V_{Ae}^2 - \omega^2)} \approx \frac{\kappaup_{in}^2 R^2}{2} \hspace*{+.2cm} \left[-\ln(\frac{\kappaup_{out}\hspace*{+.05cm}R}{2})-\gamma_e\right] \approx -\frac{\kappaup_{in}^2 R^2}{2} \hspace*{+.05cm} \ln(k\hspace*{+.05cm}R)\hspace*{+.05cm}.  \label{eq:18}
\end{equation}

Here we have taken the long-wavelength limit expansion of the Bessel $K$-function, where $\abs{\ln(kR)} \gg 1$.
As evident from the simplified dispersion relation (Eq.~(\ref{eq:18})), as $k$ approaches zero, $\kappaup_{in}^2$ diminishes much faster than $\ln(k\hspace*{+.03cm}R)$ approaches infinity. Consequently, the dispersion relation only yields a solution at the frequency equal to the internal Alfv\'en frequency ($\omega_{Ai}$), indicating that in the limit of small longitudinal wavenumbers, the dispersion relation has no solutions for frequencies exceeding the internal Alfv\'en frequency associated with fast sausage modes. Consequently, the dispersion relation, Eq.~(\ref{eq:17}), yields a solution at the frequency corresponding to the internal Alfv\'en frequency ($\omega_{Ai}$), indicating that, in the limit of small longitudinal wavenumbers, the dispersion relation lacks solutions for frequencies exceeding the internal Alfv\'en frequency associated with long wavelength fast sausage modes. Based on analytical results and referring to the numerical solution (Fig.~\ref{figure3}, solid lines), the cut-off occurring in the range of small longitudinal wavenumbers and frequencies between internal and external Alfv\'en frequencies is a phenomenon more pronounced than the ability to find solutions for frequencies even under the influence of thermal misbalance on the fast asymmetric MHD modes, with the cut-off effect remaining dominant. Therefore, in this limit, it is impossible to find an analytical expression describing long wavelength fast sausage modes.

Moreover, as evident from the numerical solution in Fig.~\ref{figure3}(a) (solid line), thermal misbalance exhibits no impact on the position of cut-off wavenumber and minimal impact on the phase speed of fast sausage modes. In Fig.~\ref{figure3}(b) (solid line), the damping rate closely approaches that of the fast kink mode (dashed line) after the cut-off wavenumber, with damping rate values being negligible, thus insignificantly influencing the overall behaviour of fast sausage modes, contrary to the slow sausage modes studied in the next subsection.

\subsection{Slow sausage modes }

\begin{figure*}
	\centering\includegraphics[width=.32\linewidth]{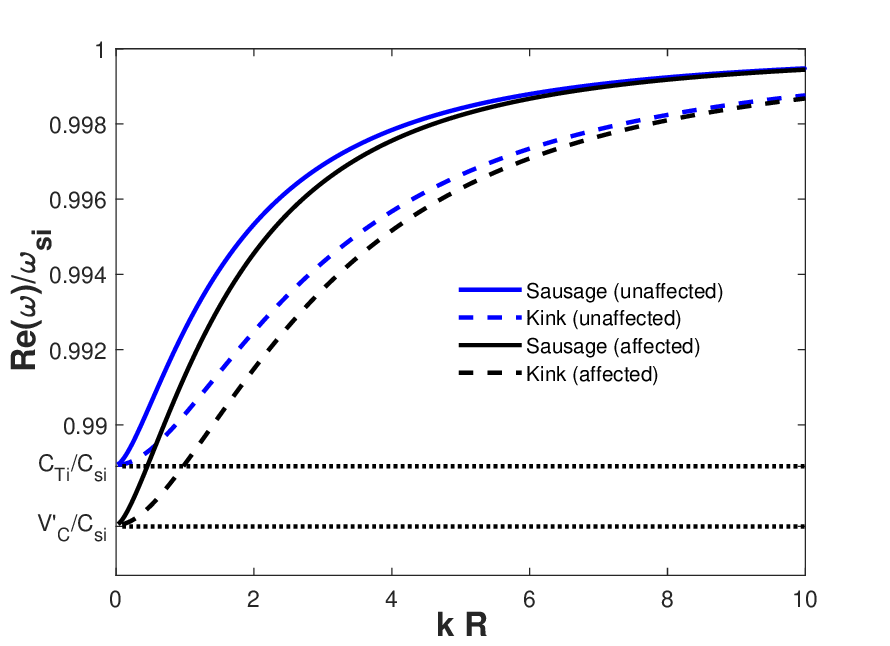}
	\centering\small  (a)
	\hfil
	\centering\includegraphics[width=.32\linewidth]{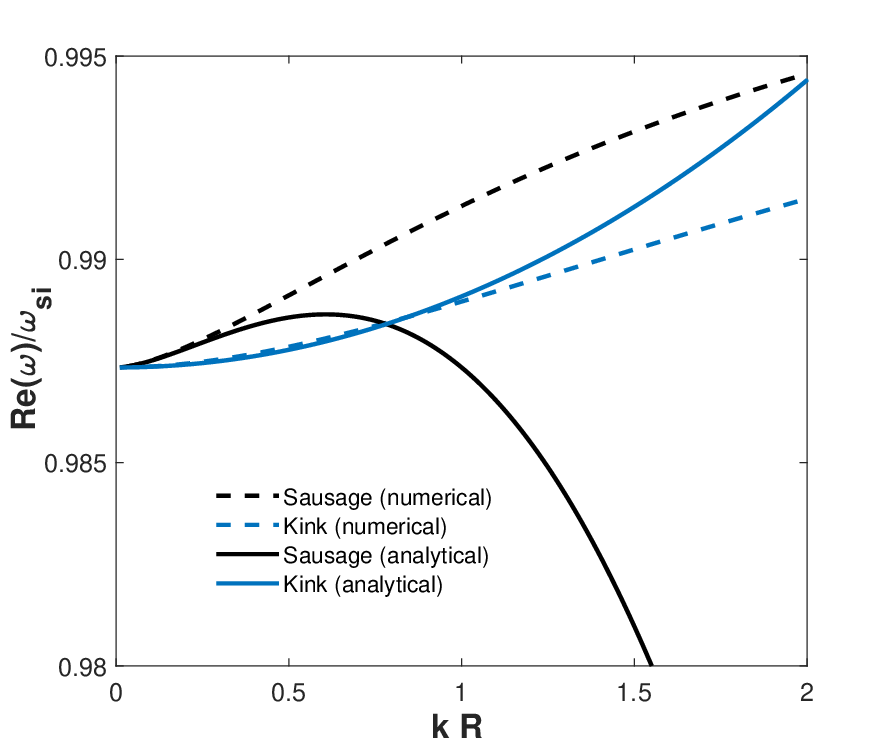}
	\centering\small  (d)
	\vfil
	\centering\includegraphics[width=.32\linewidth]{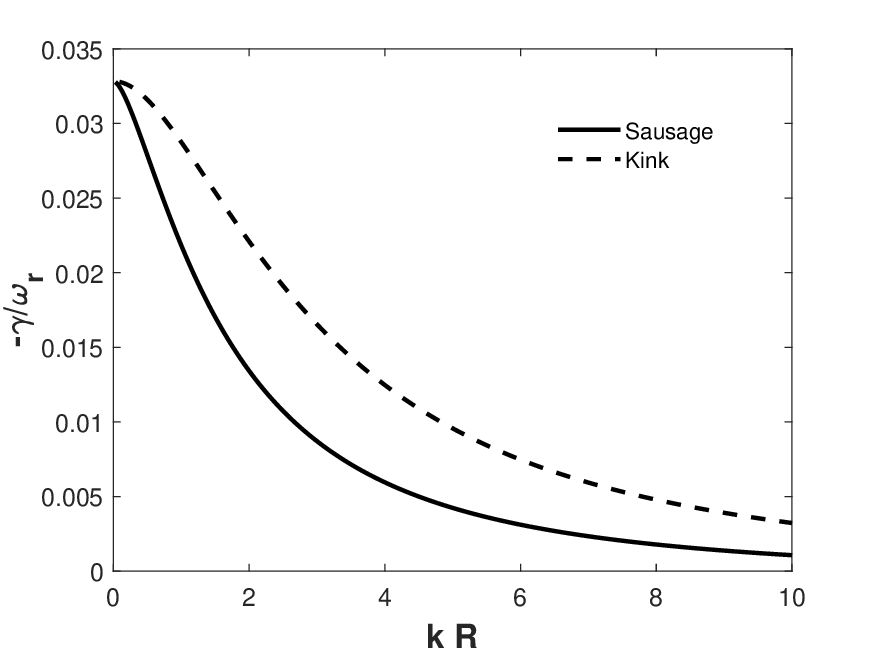}
	\centering\small (b) 
	\hfil
	\centering\includegraphics[width=.32\linewidth]{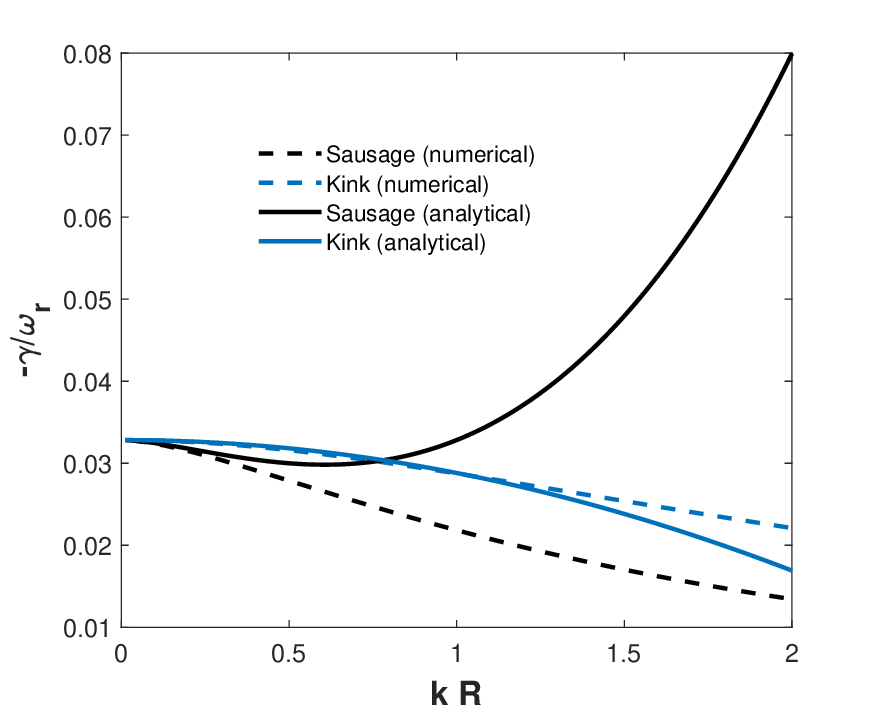}
	\centering\small (e)
	\vfil
	\centering\includegraphics[width=.32\linewidth]{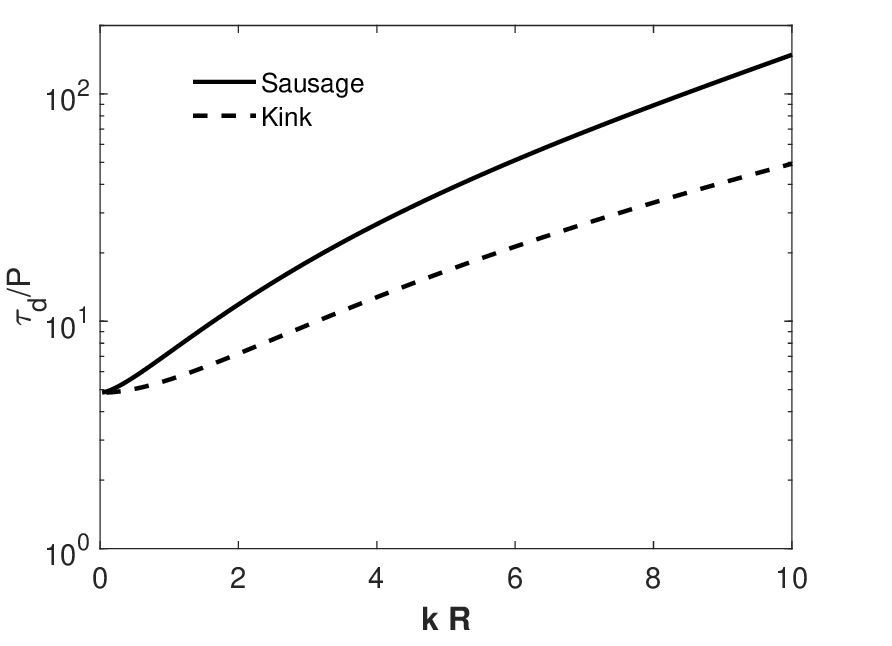}
	\centering\small (c) 
	\hfil
	\centering\includegraphics[width=.32\linewidth]{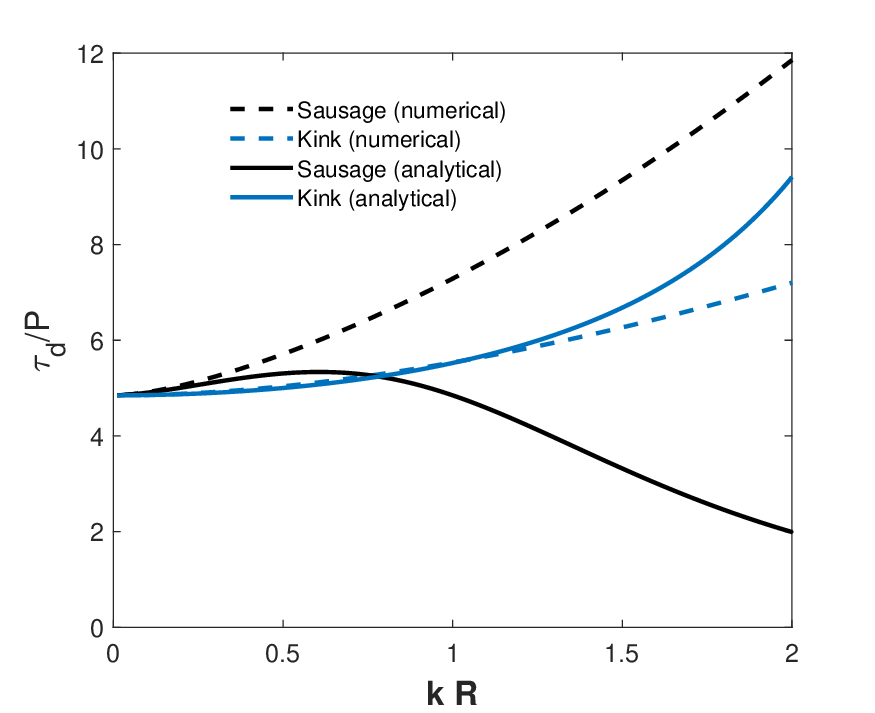}
	\centering\small (f)
	\caption{In the coronal condition and $\bar{\tau_1}=6.8$, $\bar{\tau_2}=2.4$: (a) numerical solution of the dispersion relation Eq.~(\ref{eq:16}), computed by MATLAB, for the frequency (phase speed) normalized by the internal acoustic frequency (acoustic speed) ($\mathfrak{Re}(\omega)/\omega_{si} \equiv v_{_{ph}}/C_{si}$) of the fundamental slow sausage mode without consideration of the thermal misbalance effect (unaffected) (solid blue line) and with consideration of the thermal misbalance effect (affected) (solid black line), as well as the fundamental slow kink mode unaffected (dashed blue line) and affected (dashed black line) by the thermal misbalance versus the normalized longitudinal wavenumber ($k\hspace*{+.05cm}R$). (b) Damping rate to the frequency ratio ($-\gamma/\omega_{r}$) of fundamental states of the affected slow sausage (solid line) and kink (dashed line) modes by the thermal misbalance versus the normalized longitudinal wavenumber ($k\hspace*{+.05cm}R$). (c) Damping time to the wave period ratio ($\tau_d/P$) of the affected slow sausage (solid line) and kink (dashed line) modes by the thermal misbalance versus the normalized longitudinal wavenumber ($k\hspace*{+.03cm}R$). For comparison, the analytical results obtained by Eqs.~(\ref{eq:23},~\ref{eq:29}) are shown by the black and blue solid curves in panels d, e, f.}
	\label{figure4}
\end{figure*}

Slow MHD modes arise from the analysis of wave behaviour approaching either the subsonic or sub-Alfv\'enic frequencies around the internal cusp frequency, denoted as $\omega_{_{C}}$. When $k^2 C_{T}^2 - \omega^2 = \epsilon$,  and $\epsilon$ is a very small value, the interior radial wavenumber Eq.~(\ref{eq:13}) displays a singularity implying a resonance condition in the absence of the thermal misbalance effect. At this point, the cusp frequency is equal to the internal tube frequency ($\omega_{_{Ti}}= k\hspace*{+.05cm}C_{Ti}$), see Fig.~\ref{figure4}(a) blue solid (slow sausage) and blue dashed (slow kink) lines. However, the presence of thermal misbalance introduces a slight shift in the phase speed values deviated from the values obtained without considering the influence of thermal misbalance. This results in a complex-valued cusp frequency that differs from the tube frequency in its real part, while also introducing an imaginary part directly associated with one of the characteristic timescales ($\bar{\tau_2}$) connected to the thermal misbalance effect (e.g. Fig.~\ref{figure4}(a) black solid and dashed lines). Subsequently, the cusp frequency undergoes modification and can be obtained by examining the denominator of the interior radial wavenumber given by Eq.~(\ref{eq:13}) for frequency values that cause it to approach zero. In this study, the modified cusp frequency is denoted by $\omega'_{_{C}}$ and expressed as    

\begin{equation}
	{\omega'_{_{C}}}^2 = \omega_{_{T}}^2 \left[1 + \frac{i }{2 \pi \gamma \bar{\tau_2}} \frac{C_{s}^2 + \gamma V_{A}^2}{C_{s}^2 + V_{A}^2}\right]^{-1}\hspace*{+.1cm},   \label{eq:19}
\end{equation}

As an illustration, when considering a typical condition associated with coronal active regions and normalizing all system speeds by the internal sound speed, the modified cusp speed ($V'_{_{C}}$) takes on a complex form, approximately $0.987339 - 0.0324i$ for timescales specified by values of $\bar{\tau_1}=6.8, \bar{\tau_2}=2.4$, whereas the interior tube speed ($C_{T}$) under the same conditions is approximately $0.988936$. Consequently, the recalibration of the interior radial wavenumber ($\kappaup_{in}$) with the utilization of this modified cusp frequency is denoted by $\kappaup'_{in}$ and rewritten as

\begin{equation}
	{\kappaup'_{in}}^2 = -\frac{(\omega_{s}^2 - \omega^2)(\omega_{A}^2 - \omega^2) - \frac{i \omega^2}{2 \pi \bar{\tau_2}}(\frac{\bar{\tau_2}}{\bar{\tau_1}} \omega_{s}^2 - \omega^2)}{({\omega'_{_{C}}}^2 - \omega^2)\left[C_{s}^2 + V_{A}^2 + \frac{i}{2 \pi \gamma \bar{\tau_2}}(C_{s}^2 + \gamma V_{A}^2)\right]}\hspace*{+.1cm},   \label{eq:20}
\end{equation}
where $\omega_{s}^2 \equiv k^2 C_{s}^2$ and  $\omega_{A}^2 \equiv k^2 C_{A}^2$.

To obtain the solution of the dispersion relation Eq.~(\ref{eq:17}) concerning frequencies in the vicinity of the modified cusp frequency given in Eq.~(\ref{eq:19}) under the long-wavelength limit ($k\hspace*{+.05cm}R \rightarrow 0$), we can employ a frequency expansion with the form of $\omega^2\approx {\omega'_{_{C}}}^2\left[1+\delta_{_{0}}\hspace*{+.05cm}k^2 R^2 \ln(k\hspace*{+.05cm}R)\right]$ as derived by \citet{Moreels}. This expansion is applied on the case of $m=0$, where $\delta_{_{0}}$ represents a small deviation determined by the solution of the dispersion relation around the modified cusp frequency. Therefore, utilizing the frequency expansion and expressing in term of phase speeds, Eq.~(\ref{eq:20}) transforms to

\begin{align}
	&{\kappaup'_{0in}}^2 R^2= \frac{(C_{s}^2 - {V'_{_{C}}}^2)(V_{A}^2 - {V'_{_{C}}}^2) - \frac{i {V'_{_{C}}}^2}{2 \pi \bar{\tau_2}}(\frac{\bar{\tau_2}}{\bar{\tau_1}} C_{s}^2 - {V'_{_{C}}}^2)}{ \delta_{_{0}} {V'_{_{C}}}^2 \ln(k\hspace*{+.05cm}R)\left[C_{s}^2 + V_{A}^2 + \frac{i}{2 \pi \gamma \bar{\tau_2}}(C_{s}^2 + \gamma V_{A}^2)\right]}\hspace*{+.1cm}, \label{eq:21} \\ &{V'_{_{C}}}^2=\frac{{\omega'_{_{C}}}^2}{k^2}\hspace*{+.05cm}. \nonumber   
\end{align}
where $V'_{_{C}}$ is modified cusp speed. By substituting this expression into the dispersion relation Eq.~(\ref{eq:17}) and applying standard approximations for Bessel functions, as outlined in deriving Eq.~(\ref{eq:18}), we obtain

\begin{equation}
	D_{0}(V'_{_{C}}) = (V_{A}^2 - {V'_{_{C}}}^2) + \frac{\rho_{e}}{\rho_{_0} } (V_{Ae}^2 - {V'_{_{C}}}^2) \frac{{\kappaup'_{0in}}^2 R^2 \ln(k\hspace*{+.05cm}R)}{2} = 0\hspace*{+.05cm},   \label{eq:22}
\end{equation}

By utilizing this expression and employing the frequency expansion used for slow sausage modes, the damping rate is expressed as

\begin{equation}
	\gamma_{_{damp}} = \mathfrak{Im} \left[ {\omega'_{_{C}}}\left( 1+\frac{1}{2}\delta_{_{0}} \hspace*{+.1cm} k^2 R^2 \ln(k\hspace*{+.05cm}R) \right) \right]\hspace*{+.05cm},   \label{eq:23}
\end{equation}

Here, $\delta_{_{0}}$ is obtained by solving Eq.~(\ref{eq:22}) and takes on a mathematical form of
\begin{align}
	&\delta_{_{0}} = -\frac{\rho_{e}}{2 \rho_{0} {V'_{_{C}}}^2} \left(\frac{V_{Ae}^2 - {V'_{_{C}}}^2}{V_{A}^2 - {V'_{_{C}}}^2}\right) \nonumber\\
	&\qquad \qquad \times \left[\frac{(C_{s}^2 - {V'_{_{C}}}^2)(V_{A}^2 - {V'_{_{C}}}^2) - \frac{i {V'_{_{C}}}^2}{2 \pi \bar{\tau_2}}(\frac{\bar{\tau_2}}{\bar{\tau_1}} C_{s}^2 - {V'_{_{C}}}^2)}{C_{s}^2 + V_{A}^2 + \frac{i}{2 \pi \gamma \bar{\tau_2}}(C_{s}^2 + \gamma V_{A}^2)}\right]\hspace*{+.05cm}.   \label{eq:24}
\end{align}

In Figs.~\ref{figure4}(d-f), numerical solutions (dashed black lines) obtained by numerically solving Eq.\ref{eq:16} for the axisymmetric MHD modes with $m=0$ are compared with analytical solutions (solid black lines) derived from the frequency expansion $\omega^2\approx {\omega'_{_{C}}}^2\left[1+\delta_{_{0}} k^2 R^2 \ln(kR) \right]$, where the parameters are obtained from Eqs.~(\ref{eq:19}, \ref{eq:24}), and Eq.~(\ref{eq:23}) for the fundamental state of slow sausage modes. The comparison reveals a high degree of consistency between the numerical and analytical solutions, particularly evident in the long-wavelength limit (small longitudinal wavenumbers $k$), affirming the validity of our analytical expressions.

\subsection{Fast kink modes}

Higher azimuthal harmonics ($m>0$) of the dispersion relation in Eq.~(\ref{eq:16}) can take the form of

\begin{align}
	&D_{m}(\omega) = (k^2 V_{A}^2 - \omega^2) \nonumber \\
	&\quad -\frac{\rho_{e}}{\rho_{_0}} \frac{\kappaup_{in}}{\kappaup_{out}} (k^2 V_{Ae}^2 - \omega^2) \frac{\left(\frac{m}{\kappaup_{in} R} J_{m}(\kappaup_{in} R)- J_{m+1}(\kappaup_{in} R)\right) K_{m}(\kappaup_{out} R)}{J_{m}(\kappaup_{in} R) \left(\frac{m}{\kappaup_{out} R} K_{m}(\kappaup_{out} R)- K_{m+1}(\kappaup_{out} R)\right)} \nonumber \\
	&\qquad\qquad = 0\hspace*{+.05cm}, \hspace*{+.9cm} m>0\hspace*{+.05cm}. \label{eq:31}
\end{align}

In the limit of long wavelength, the Bessel approximations for small values of the longitudinal wavenumber ($k$) allow us to rewrite Eq.~(\ref{eq:31}) to a reduced form of

\begin{equation}
	\frac{\rho_{_0} (k^2 V_{A}^2 - \omega^2)}{\rho_{e} (k^2 V_{Ae}^2 - \omega^2)} \approx \frac{\kappaup_{in}^2 R^2}{2m (m+1)}-1\hspace*{+.2cm}, \hspace*{+.9cm} m>0\hspace*{+.05cm}.  \label{eq:25}
\end{equation} 

Subsequently, in the regime of small longitudinal wavenumbers where the interior radial wavenumber Eq.~(\ref{eq:13}) approaches zero, for non-zero azimuthal harmonic numbers (including $m=1$ to describe kink modes), the following expression can be obtained from Eq.~(\ref{eq:25}):

\begin{equation}
	\frac{\rho_{_0} (k^2 V_{A}^2 - \omega^2)}{\rho_{e} (k^2 V_{Ae}^2 - \omega^2)} \approx -1\hspace*{+.05cm}, \label{eq:26}
\end{equation} 

which is only solved by the kink frequency of 

\begin{equation*}
	\omega_{k}^2 = \frac{\rho_{_0} V_{A}^2 + \rho_e V_{Ae}^2}{\rho_{0} + \rho_{e}} \hspace*{+.2cm} k^2\hspace*{+.05cm}. 
\end{equation*} 

It can be readily inferred from Eq.~(\ref{eq:25}) that thermal misbalance has no significant effect on fast either kink or higher azimuthal harmonic waves ($m>1$) in the long wavelength limit. The numerical analysis is also shown in Fig.~\ref{figure3} (dashed lines) for fast kink modes that thermal misbalance has no impact on the fundamental fast kink frequency (phase speed). However, a very small decrement is observable for the damping rate in Fig.~\ref{figure3} (b) (dashed black line) for small values of longitudinal wavenumbers which may be inferred as suppressed damping state of kink oscillations \citep{Agapova}, and very interesting for further research. In addition, a slight increase is observed for greater values of longitudinal wavenumbers while the damping rate values are very small and negligible. The conclusion of no significant impact is totally different for slow kink modes. 

\subsection{Slow kink modes} 

In the case of slow kink modes, the criterion for finding an analytical expression is similar to what was done in subsection 3.2 for slow sausage modes. The analytical expression for the damping rate is derived by employing the modified cusp frequency ($\omega'_{_{C}}$) defined in Eq.~(\ref{eq:19}) and the rewritten form of the interior radial wavenumber ($\kappaup'_{in}$) in Eq.~(\ref{eq:20}). By setting the dispersion relation Eq.~(\ref{eq:31}) for the first azimuthal wavenumber ($m=1$), and employing the Bessel function approximations under the assumption of $k\hspace*{+.05cm}R\ll1$, an approximate form of the dispersion relation Eq.~(\ref{eq:31}) describing the long wavelength slow kink modes is obtained as follows

\begin{equation}
	D_{1}(V_{ph}) = (V_{A}^2 - V_{ph}^2) +\frac{\rho_{e}}{\rho_{_0} } (V_{Ae}^2 - V_{ph}^2) \left[1-\frac{{\kappaup'_{in}}^2 R^2}{4}\right] = 0\hspace*{+.05cm}.   \label{eq:27}
\end{equation}
where $V_{ph}=\frac{\omega}{k}$. Once again, similar to the slow sausage modes, to analytically study the kink wave around the cusp frequency in the long-wavelength limit, we need to employ a frequency expansion. However, this time, it is specified for the case of $m=1$ with the form of $\omega^2\approx {\omega'_{_{C}}}^2(1+\delta_{_{1}} k^2 R^2)$ ~\citep{EandR}, where the small deviation $\delta_{_{1}}$ is obtained by solving the dispersion relation Eq.~(\ref{eq:27}) for the modified cusp speed ($V'_{_{C}}$).

\begin{figure*}
	\centering\includegraphics[width=0.32\linewidth]{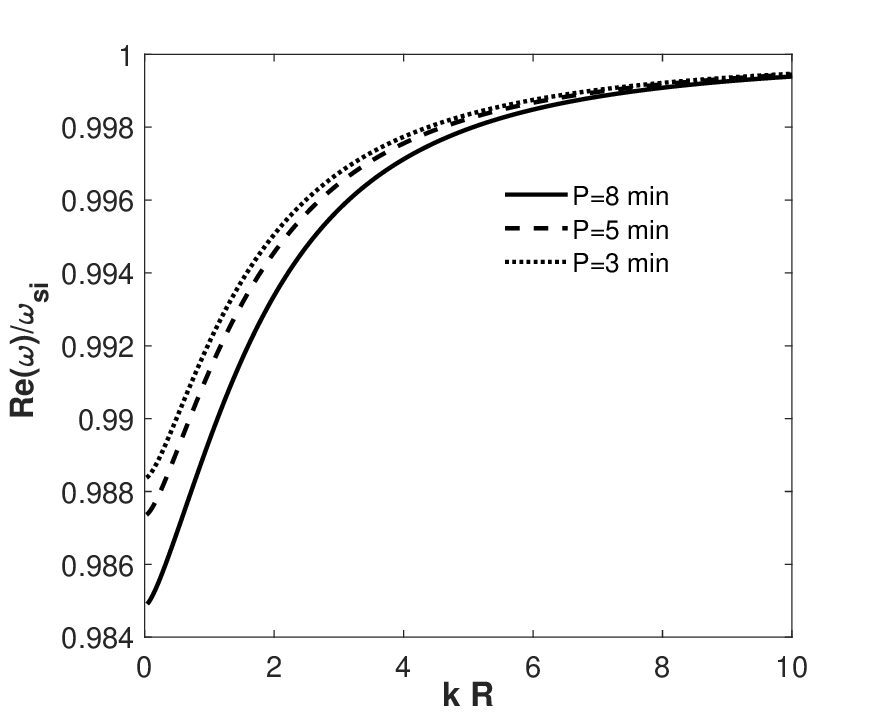}
	\centering\small  (a)
	\hfil
	\centering\includegraphics[width=0.34\linewidth]{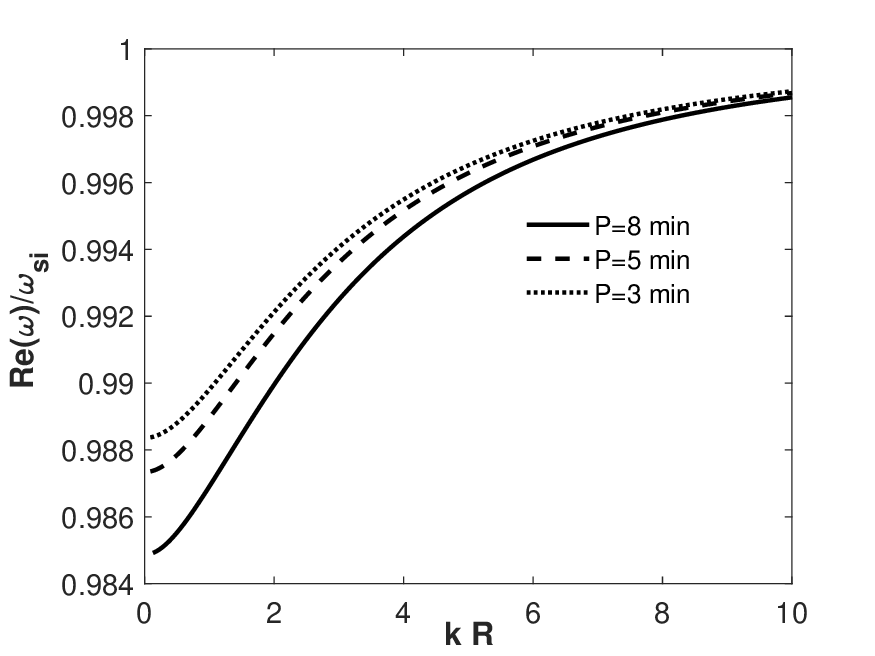}
	\centering\small  (d)
	\vfil
	\centering\includegraphics[width=0.32\linewidth]{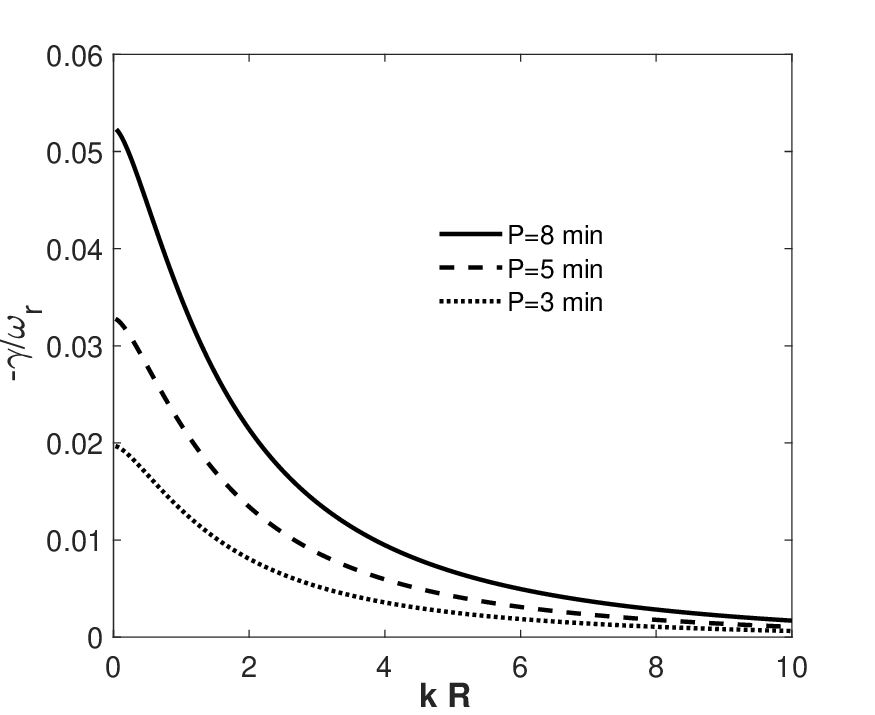}
	\centering\small (b) 
	\hfil
	\centering\includegraphics[width=0.32\linewidth]{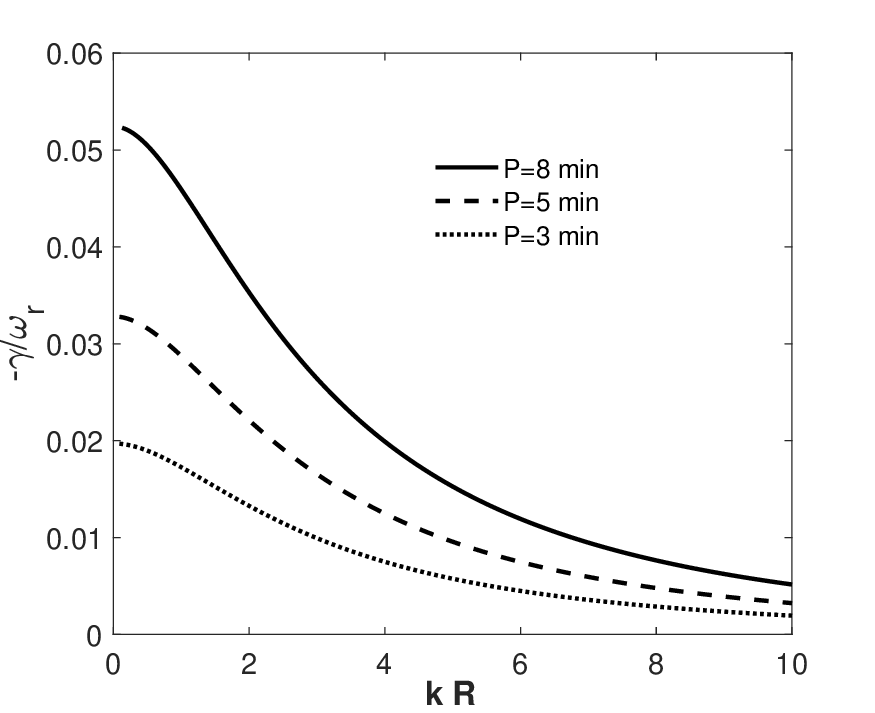}
	\centering\small (e)
	\vfil
	\centering\includegraphics[width=0.32\linewidth]{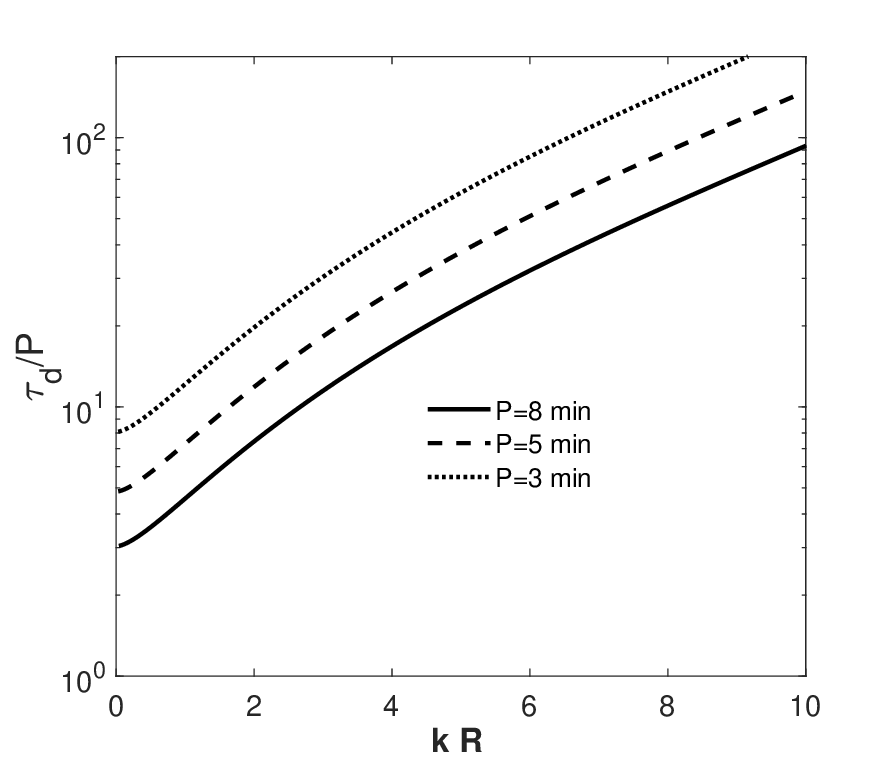}
	\centering\small (c) 
	\hfil
	\centering\includegraphics[width=0.32\linewidth]{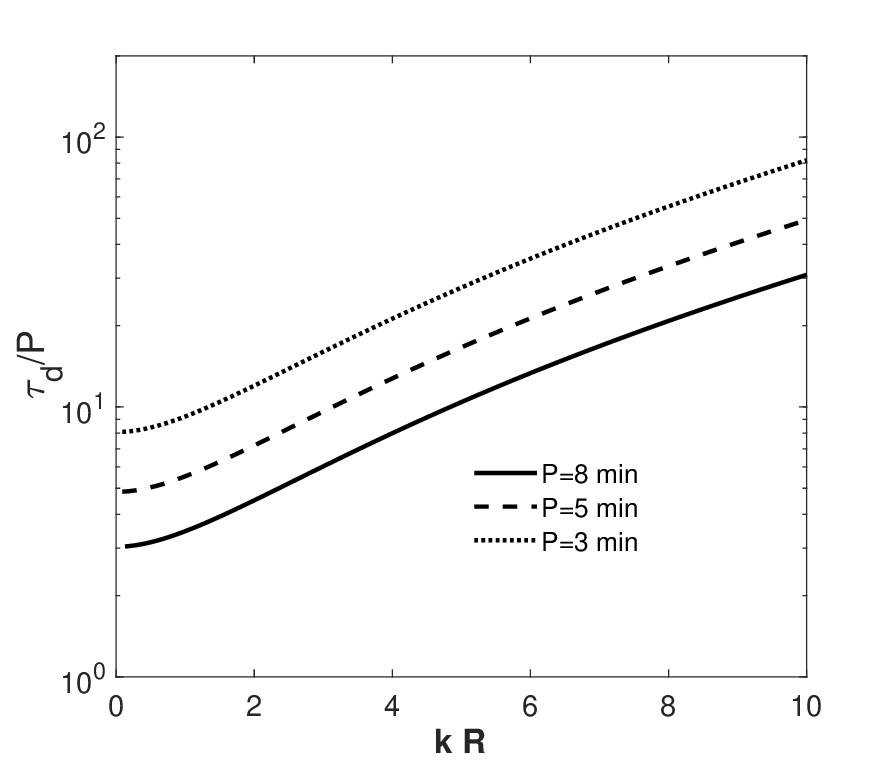}
	\centering\small (f)
	\caption{In the coronal condition, the figure displays: (a), (b), and (c) the normalized numerical solutions of the dispersion relation Eq.~(\ref{eq:17}) by the internal acoustic speed ($C_{s}$) for sausage modes ($m=0$) with different periods of 8 min ($\bar{\tau_1}=4.25$, $\bar{\tau_2}=1.5$) (solid line), 5 min ($\bar{\tau_1}=6.8$, $\bar{\tau_2}=2.4$) (dashed line), and 3 min ($\bar{\tau_1}=11.34$, $\bar{\tau_2}=4$) (dotted line); and (d), (e), and (f): the normalized numerical solutions of the dispersion relation Eq.~(\ref{eq:31}) by the internal acoustic speed ($C_{s}$) for kink modes ($m=1$) with different periods of 8 min ($\bar{\tau_1}=4.25$, $\bar{\tau_2}=1.5$) (solid line), 5 min ($\bar{\tau_1}=6.8$, $\bar{\tau_2}=2.4$) (dashed line), and 3 min ($\bar{\tau_1}=11.34$, $\bar{\tau_2}=4$) (dotted line).}
	\label{figure6}
\end{figure*}

By utilizing the frequency expansion, the radial wavenumber $\kappaup'_{in}$ given by Eq.~(\ref{eq:20}) for the modified cusp speed has the following mathematical expression

\begin{equation}
	{\kappaup'_{1in}}^2 R^2= \frac{(C_{s}^2 - {V'_{_{C}}}^2)(V_{A}^2 - {V'_{_{C}}}^2) - \frac{i {V'_{_{C}}}^2}{2 \pi \bar{\tau_2}}(\frac{\bar{\tau_2}}{\bar{\tau_1}} C_{s}^2 - {V'_{_{C}}}^2)}{\delta_{_{1}} {V'_{_{C}}}^2 \left(C_{s}^2 + V_{A}^2 + \frac{i}{2 \pi \gamma \bar{\tau_2}}(C_{s}^2 + \gamma V_{A}^2)\right)}\hspace*{+.05cm},   \label{eq:28}
\end{equation}

By replacing this expression into the dispersion relation Eq.~(\ref{eq:27}) instead of ${\kappaup'_{in}}^2 R^2$, for the phase speed equal to $V'_{_{C}}$, and solving it for the deviation of $\delta_{_{1}}$, an analytical expression for the damping rate is obtained as follows

\begin{equation}
	\gamma_{_{damp}} = \mathfrak{Im}\left[{\omega'_{_{C}}}\left(1 + \frac{1}{2} \delta_{_{1}} k^2 R^2\right)\right]\hspace*{+.05cm},   \label{eq:29}
\end{equation}

Where includes the deviation of $\delta_{_{1}}$ defining with the following expression

\begin{align}
	&\delta_{_{1}} = \left(\frac{(C_{s}^2 - {V'_{_{C}}}^2)(V_{A}^2 - {V'_{_{C}}}^2) - \frac{i {V'_{_{C}}}^2}{2 \pi \bar{\tau_2}}(\frac{\bar{\tau_2}}{\bar{\tau_1}} C_{s}^2 - {V'_{_{C}}}^2)}{4 {V'_{_{C}}}^2 \left[C_{s}^2 + V_{A}^2 + \frac{i}{2 \pi \gamma \bar{\tau_2}}(C_{s}^2 + \gamma V_{A}^2)\right]}\right)\nonumber \\
	&\qquad\qquad \times\left[1+\frac{\rho_{0}}{ \rho_{e}} \frac{V_{A}^2 - {V'_{_{C}}}^2}{V_{Ae}^2 - {V'_{_{C}}}^2}\right]^{-1}.   \label{eq:30}
\end{align}

The comparison in Figs.~\ref{figure4}(d-f) reveals a high degree of consistency between the numerical (dashed blue lines) and analytical (solid blue lines) solutions for the fundamental state of slow kink modes, arisen from Eq.~(\ref{eq:31}) (for transverse harmonic m=1) and Eq.~(\ref{eq:30}) respectively, particularly evident in the long wavelength limit (small longitudinal wavenumbers $k$), and slight deviation observed for larger values of longitudinal wavenumbers, affirming the validity of our analytical equation (Eq.~(\ref{eq:29})).

Finally, in Fig.\ref{figure6}, we numerically analyse the behaviour of slow sausage and kink waves and their associated damping characteristics across varying periods of 3, 5, and 8 minutes. It is observed that longer periods correspond to smaller values of $\bar{\tau_1}$ and $\bar{\tau_2}$, leading to enhanced damping rates. The analytical expressions in Eqs.~(\ref{eq:23}, \ref{eq:29}) also demonstrate this conclusion.

\section{Conclusions}
\begin{figure}
	\centering\includegraphics[width=0.7\linewidth]{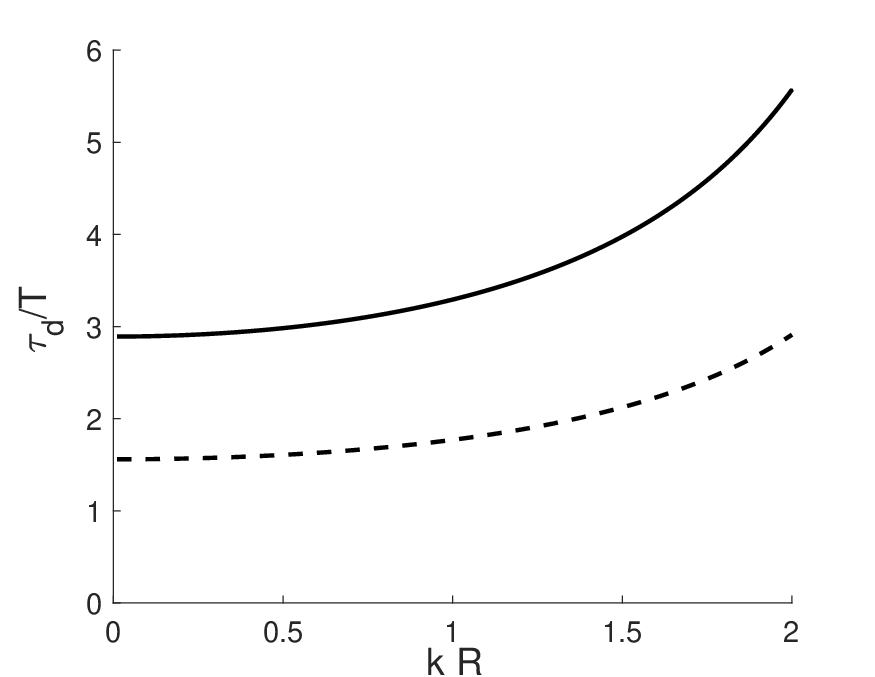}
	
	\caption{In the coronal condition, for temperatures associated with SUMER oscillations with periods of 15.7 minutes at 6.3 MK ($\bar{\tau_1}\approx2.36$, $\bar{\tau_2}\approx0.76$) (dashed line) and 13.3 minutes at 8.9 MK ($\bar{\tau_1}\approx4.89$, $\bar{\tau_2}\approx1.43$) (solid line), as calculated using the analytical expression in Eq.~(\ref{eq:29}), the figure depicts the ratio of the damping time to the wave period ($\tau_{d}/P$) of the affected slow transverse MA modes by the thermal misbalance versus $k\hspace*{+.03cm}R$.}
	\label{figure7}
\end{figure}

In this study, we investigated the propagation of MHD waves dispersed in a cylindrical waveguide and affected by non-adiabatic factors, with thermal misbalance at the centre of focus. We characterized the impact of thermal misbalance resulting from the deviation from thermal equilibrium in the coronal region between heating and radiative cooling mechanisms, using two distinct dimensionless characteristic times $\bar{\tau_1}$ and $\bar{\tau_2}$. These times play a crucial role in either attenuating or amplifying MHD wave amplitudes within a cylindrical magnetic waveguide in a coronal uniform plasma. The characteristic times are directly linked to the heating-cooling function $\emph{Q}(\rho,T)$, the sole non-adiabatic term considered in this analysis. We deliberately selected positive values for these timescales, ensuring that $\bar{\tau_1}-\bar{\tau_2}>0$, aligning with constraints consistent with the stronger damping (enhanced damping) regimes identified in \citet{Kolotkov2020}. The ensuing results are presented as follows
\begin{enumerate}
	\item The dispersion relation governing MHD modes in cylindrical waveguides immersed in a uniform finite magnetic field and subjected to the discrepancy between heating and radiative cooling mechanisms was obtained. The impact of thermal misbalance, as elucidated by two characteristic times linked to the derivative of the heating-cooling function with respect to density perturbation in constant temperature and with respect to temperature perturbation in constant density, initially introduced by \citet{Kolotkov2019}, were normalized by the MHD wave period. These normalized characteristic timescales are employed to encapsulate the influence of thermal misbalance properties on the radial wavenumber. 
	
	\item Both the analytical and numerical solutions in the limit of fast MHD modes were obtained to show that thermal misbalance exhibits only a marginal impact on these modes which stems from the consideration that, in high-frequency regimes ($\omega \gg 1/\min\{|\tau_1|,|\tau_2|\}$) \citep{Nakariakov2017}, the influence of thermal misbalance on the damping rate remains minimal and negligible, in consistency with the assumption established firstly by \citet{Zavershinskii}. However, the slight decrement observed in the small longitudinal wavenumbers for the fundamental fast kink wave, Fig.~\ref{figure3}(b) (dashed line), may signify a suppressed damping mode, which merits further investigation.
		
	\item In the case of slow MHD modes within a magnetic tube, both analytical and numerical results were obtained. Analytical expressions in Eqs.~(\ref{eq:23}, ~\ref{eq:29}) governing the damping rates in the long wavelength limit were derived for slow sausage and kink modes respectively, demonstrating exemplary agreement with the corresponding numerical results presented in Fig.~\ref{figure4}. The equations and figures collectively reveal that thermal misbalance significantly impacts slow modes in comparison to fast modes.
	Furthermore, our analysis indicates that the influence of thermal misbalance on slow MHD wave propagation in the coronal conditions is more pronounced in asymmetric modes than in axisymmetric ones. This result opposes with the expected behaviour, as MA waves characterized by higher compressibility exhibit greater sensitivity to thermal misbalance. Figures~\ref{figure4}(b)\&(e) illustrate that the damping rates for slow kink modes are higher than those for slow sausage modes with the same period. This finding suggests that thermal misbalance has a more substantial impact on slow asymmetric MA modes than their axisymmetric counterparts. In fast modes, as depicted in Fig.~\ref{figure3}, a slight inverse relationship is evident. As a result, we can conclude that the MA waves owning the faster phase speeds are less affected than the slower ones by thermal misbalance.
	
	\item In the investigation of either high-frequency (small period) waves or elevated temperatures, an examination of the surface plot Fig.~\ref{figure2} reveals that both normalized characteristic timescales, denoted as $\bar{\tau_1}$ and $\bar{\tau_2}$, tend towards their maximum values. This observation suggests that, under the specified conditions which are subject to the heating model and consequently the values considered for the parameters, $a$ and $b$, in the heating function Eq.~(\ref{eq:6}), the influence of thermal misbalance on the dynamical behaviour of cylindrical MHD modes is minimized. This is because the characteristic timescales, in their respective maximum values, minimize the imaginary part of the radial wavenumbers, as evident from Eqs.~(\ref{eq:13} and \ref{eq:15}). On the contrary, the diminution of the characteristic timescales exerts a pronounced influence, leading to an enhanced damping rate for MHD waves, particularly for slow modes. Moreover, as $\bar{\tau_1}$ and $\bar{\tau_2}$ exhibit a correlation with the temperature of the active plasma region, as observable in Fig.~\ref{figure2}, the damping rates Eqs.~(\ref{eq:23}, \ref{eq:29}) consequently demonstrate a correlated behaviour with temperature. In such a way, the damping rate decreases with rising temperature and increases as the temperature decreases to the 1-5 MK range which is evident in Fig.~\ref{figure7} for transverse slow oscillations as well. 
	
	Importantly, our analysis underscores the significance of $\bar{\tau_2}$ over $\bar{\tau_1}$. Especially, the heating-cooling function's rate with respect to temperature, encapsulated in $\bar{\tau_2}$, plays a more prominent role in shaping wave dynamics. This is particularly evident as $\bar{\tau_2}$ is the exclusive timescale in the denominator of the fourth power of wave frequency in the imaginary parts of the radial wavenumbers (Eqs.~(\ref{eq:13}) and (\ref{eq:15})), and it is the only characteristic timescale appearing in the cusp frequency Eq.~(\ref{eq:19}). Consequently, smaller values of $\bar{\tau_2}$ signify a more substantial impact of thermal misbalance.
	
	\item As mentioned, the analytical damping rates of slow MHD waves were determined in Eqs.~(\ref{eq:23}) and (\ref{eq:29}), and the obtained values were found to be comparable to their respective wave periods, aligning well with observed characteristics. To elucidate this relationship, we employed the analytical expression derived from Eq.~(\ref{eq:29}), which describes the damping rate of slow transverse MA modes in a magnetic tube. The results, as illustrated in Fig.~\ref{figure7}, demonstrate that the fraction of damping time per the period of the wave for SUMER oscillations falls within the range of approximately 1.55 to 2.85 times the wave period at 6.3 MK and approximately 2.9 to 5.5 times the wave period at 8.9 MK, for a range of normalized longitudinal wavenumbers ($k\hspace*{+.03cm}R$) between 0 and 2. This finding is consistent with the observed rapid damping of slow modes, especially in the limit of long-wavelength waves, substantiating the validity of the analytical model in capturing the damping behaviour of the studied MHD waves. 
	
	\item Adjusting the thermal misbalance characteristic timescales, $\bar{\tau_1}$ and $\bar{\tau_2}$, to match the observed ratio of rapidly damping or amplifying times per the period of the oscillations in various layers of the solar atmosphere offers seismological insights on plasma characteristics or mechanisms, for example, plasma density or the heating function. It is also examinable alongside other methods, especially in the context of identifying the heating function. As an instance, in this study, we demonstrated that the selected values for these timescales, taken from the work of \citet{Kolotkov2019}, show compatibility with the damping times observed for the considered SUMER oscillations, as shown in Fig.~\ref{figure7}. This study propose that gathering the values of these timescales consistent with each oscillation provides a set of statistical data to specify the heating function form with the more accurate values of the heating function exponents of $a$ and $b$ in Eq.~(\ref{eq:6}). 
\end{enumerate}

In conclusion, our current study has provided robust analytical and numerical evaluations of MHD modes in a cylindrical configuration influenced by thermal misbalance. Building upon this foundation, the next crucial steps involve extending our theoretical framework to incorporate other non-adiabatic terms, notably thermal conduction and scalar resistivity. Given that thermal conduction always acts as a damping for slow MA modes, as demonstrated by \citet{Duckenfield}, incorporating it alongside thermal misbalance promises more precise estimates for damping times, aligning closely with observed values.

Furthermore, the methodology and results presented in this article can be replicated for alternative sets of thermal misbalance characteristic times, shedding light on non-decaying, amplifying, and suppressed damping oscillations, aligning with the states identified by \citet{Kolotkov2020}. It is also valuable to consider other physical parameters participating in the heating function such as magnetic field \citep{Duckenfield}, and study their effects on MHD modes inside a cylindrical waveguide. These investigations can contribute to refining heating function models governing magnetic structures within the solar atmosphere.

An intriguing avenue for future research involves introducing the concept of a resonant layer at the boundary between the inside and outside of solar tubes. Moreover, a differential operator, similar to those defined for the geometrical constraint and thermal misbalance effect in Section 2, can be introduced to describe the impact of thermal conduction as an additional cooling process alongside radiative cooling in the dispersion relation Eq.~(\ref{eq:7}) \citep{Field}, and by incorporating the characteristic timescale ($\tau_{cond}$) related to thermal conduction \citep{Kolotkov2019} in the solar atmosphere, we anticipate more accurate predictions of damping times for slow modes observed in this region.    

Expanding our focus to include the non-linear evolution of MHD waves in non-uniform plasmas, especially due to gravitational stratification, promises exciting insights. Exploring the impact of thermal misbalance on non-linear phenomena, such as shock waves, becomes particularly relevant in understanding energy release processes in the solar atmosphere, unravelling the heating mechanism. Among these, investigating the compressible aspects of non-linear evolution of Alfv\'en waves, and their response to thermal misbalance, especially when propagating through inhomogeneous plasma and forming shock waves, or how thermal misbalance influences force equilibrium, specifically its effects on the ponderomotive force, presents opportunities for further investigations.

Finally, it can be the best point to run simulations that play a pivotal role in obtaining a more tangible understanding of how non-adiabatic terms affect MHD waves within magnetic structures in the solar atmosphere. The implications derived from the findings presented in this article and the proposed follow-up investigations significantly enhance our understanding of the intricate dynamics that govern the solar corona, particularly concerning its thermal characteristics. 

\begin{acknowledgements}
	CHIANTI is a collaborative project involving George Mason University, the University of Michigan
	(USA) and the University of Cambridge (UK).
	TVD received financial support from the Flemish Government under the long-term structural Methusalem funding program, project SOUL: Stellar evolution in full glory, grant METH/24/012 at KU Leuven, the DynaSun project (number 101131534 of HORIZON-MSCA-2022-SE-01), and also a Senior Research Project (G088021N) of the FWO Vlaanderen.
	DYK acknowledges funding from the STFC consolidated grant ST/X000915/1 and Latvian Council of Science Project No. lzp2022/1-0017.     
\end{acknowledgements}

\bibliographystyle{aa}
\bibliography{references}

\begin{appendix} 
	\section{Derivation of the differential operators in Eq.~(\ref{eq:8})}
	\label{appendix:A}
	Considering the system at an initial magnetohydrodynamical equilibrium and by applying small amplitude perturbations to the governing equations (Eqs.~(\ref{eq:1}-\ref{eq:06})), their linearized forms are derived, respectively, as follows
		
\begin{align}
	&\rho_{_{0}} \frac{\partial {\bf v}_{_{1}}}{\partial t} = -\nabla p_{_{1}} - \frac{1}{4\pi} \left[\nabla (\vec{B}_{_0} \cdot \vec{B}_{_1})-(\vec{B}_{_0} \cdot \nabla) \vec{B}_{_1}\right] ,  \label{eq:32}\\
	&\frac{\partial \rho_{_{1}}}{\partial t} + \rho_{_{0}} \nabla \cdot{\bf v}_{_{1}} = 0,   \label{eq:33}\\
	&\frac{\partial \vec{B}_{_1}}{\partial t} = - \vec{B}_{_0} \nabla \cdot {\bf v}_{_{1}},   \label{eq:34}\\
	&p_{_{1}} = \frac{k_{_{B}}}{m} \left(\rho_{_{0}} T_{_1} + T_{_0} \rho_{_{1}}\right),   \label{eq:35}\\
	&\frac{1}{\gamma -1} \left(\frac{\partial p_{_{1}}}{\partial t} - \frac{\gamma p_{_{0}}}{\rho_{_{0}}} \frac{\partial \rho_{_{1}}}{\partial t} \right) = - \rho_{_{0}} \left[\rho_{_{1}} \emph{Q}_{T} + T_{_1} \emph{Q}_{\rho} \right],   \label{eq:36}
\end{align}
where, the first-order perturbed parameters are denoted by the subscript 1, while the equilibrium quantities are denoted by the subscript 0. For the definitions of the quantities $\emph{Q}_{\rho}$ and $\emph{Q}_{T}$, refer to Eq.~(\ref{eq:9}). By substituting the first order magnetic field perturbation $(\vec{B}_{_1})$ in Eq.~(\ref{eq:32}), using Eq.~(\ref{eq:34}), we obtain
\begin{equation}
	\frac{\partial}{\partial t}\nabla^2 p_{_{1}}=\rho_{_{0}} \left(V_{A}^2 \nabla^2 (\nabla \cdot{\bf v}_{_{1}}) - V_{A}^2 \frac{\partial^2}{\partial z^2} \nabla \cdot{\bf v}_{_{1}}-\frac{\partial^2}{\partial t^2} \nabla \cdot{\bf v}_{_{1}}\right),   \label{eq:37}
\end{equation}
where $V_{A} = B_{_0} (4\pi \rho_{0})^{-1/2}$ is Alfv\'en speed. The first order temperature perturbation ($T_{_1}$) is also obtained by Eq.~(\ref{eq:35}) as
\begin{equation}
	T_{_1}=\frac{1}{\rho_{_{0}}} \left(\frac{m}{k_{_{B}}} p_{_{1}} -T_{_0} \rho_{_{1}}\right),   \label{eq:38}
\end{equation}

Replacing the perturbed physical parameters in Eq.~(\ref{eq:36}) with their expressions from Eqs.~(\ref{eq:33}, \ref{eq:37}, \ref{eq:38}) and rearranging the equation by $\nabla \cdot {\bf v}_{_{1}}$, we derive the following differential equation (the subscript 1 is omitted):

\begin{align}
	&\frac{\partial^2}{\partial t^2} \left[\frac{\partial^2}{\partial t^2} (\nabla \cdot {\bf v}) - (V_{A}^2+C_{s}^2) \nabla^2 (\nabla \cdot {\bf v}) \right] + V_{A}^2 C_{s}^2 \frac{\partial^2}{\partial z^2} \nabla^2 (\nabla \cdot {\bf v}) \nonumber \\
	&\qquad = - \rho_{_0} (\gamma-1) \emph{Q}_{\rho} \frac{\partial}{\partial t} \nabla^2 (\nabla \cdot {\bf v})\nonumber \\
	&\qquad- \frac{\emph{Q}_{_{T}}}{C_{_V}}\frac{\partial}{\partial t}\left(\frac{\partial^2}{\partial t^2}(\nabla \cdot {\bf v}) - V_{A}^2 \nabla^2 (\nabla \cdot {\bf v}) + V_{A}^2 \frac{\partial^2}{\partial z^2} (\nabla \cdot {\bf v}) \right) \nonumber \\
	&\qquad +(\gamma-1) T_{_0} \emph{Q}_{_{T}} \frac{\partial}{\partial t} \nabla^2 (\nabla \cdot {\bf v})\hspace*{+.05cm}.   \label{eq:40}
\end{align}
where $C_s = \sqrt{\gamma k_{_{B}} T_{_0} m^{-1}}$ and $C_{_V} = (\gamma - 1)^{-1} k_{_{B}}/m$ are acoustic speed and specific heat capacity at constant volume, respectively.
It is interesting to note when the context of study is focused only on the plasma with lack of thermal misbalance effect, all the terms on the RHS of Eq.~(\ref{eq:40}) disappear, and the outcome aligns with Eq.~(3a) of \citet{EandR}. 

\end{appendix}

\end{document}